\documentclass[amstex,epsf,amssymb,usenatbib]{mnras}
\usepackage{epsf,graphics,subfig}
\usepackage{amsmath,amssymb,natbib}
\usepackage{lmodern}
\usepackage{cite}

\usepackage{rotating,times,graphicx,latexsym,color,longtable,lscape} 

\def\aj{AJ}                   
\def\apj{ApJ}                 
\def\apjl{ApJ}                
\def\apjs{ApJS}               
\def\aap{A\&A}                
\def\mnras{MNRAS}             
\def\pasj{PASJ}               
\def\nat{Nature}              

\title[Simulations of circumbinary wind  and YSO jet]{ The nature of a primary jet within a circumbinary disc outflow in a young stellar system}
   
\author[C. Lynch \& M.D. Smith]
{ Chris Lynch $^{1}$\thanks{E-mail: captainlockheed@btopenworld.com } \&
{Michael D. Smith $^{1}$\thanks{E-mail: m.d.smith@kent.ac.uk, }  }\\
$^{1}$Centre for Astrophysics \& Planetary Science, The University of Kent, Canterbury, Kent CT2 7NH, U.K. }                                                                                                                                                             
\date{Accepted .....
      Received ..... ;
      in original form .....}
                                                                                                                                                                  
\pagerange{\pageref{firstpage}--\pageref{lastpage}}
\pubyear{2019}

\begin{document}
                                                                                                                                                             
\maketitle
                                                                                                                                                             
\label{firstpage}
                                                                                                                                                             
\begin{abstract}
Most stars form in binaries, and both stars may grow by accreting material from a circumbinary disc onto their personal discs. 
We suspect that in many cases a wide molecular wind will envelope a collimated atomic jet emanating from close to an orbiting young star.
This so-called Circumbinary Scenario is explored here in order to find common  identifiable properties. The dynamical set up is studied with  three dimensional simulations  with chemistry and cooling included.  We extract the properties on scales of order 100\,AU 
and compare to the  Co-Orbital Scenario  in which the wind and jet sources are in orbit.
 We find that the rapid orbital motion generates a wide  ionised sheath around the jet core with a large opening angle at the base. This is independent of the presence of the surrounding molecular outflow. However, the atomic jet is recollimated beyond $\sim$ 55\,AU when the molecular outflow restricts the motion of the ambient medium which, in turn, confines the jet. These physical properties are relayed to the optical H$\alpha$ imaging, providing a means of distinguishing between models. The high excitation sheath and recollimation region can be explored on these scales through the next generation of instruments. However, in general, the amount and location of the ionised material, whether in the knots or the sheath, will depend on several parameters including the orbital period, axis alignment and pulse amplitude.
 
\noindent 
\end{abstract} 
  
\begin{keywords}
 hydrodynamics --  ISM: jets and outflows -- stars: formation -- stars: pre-main-sequence
   \end{keywords}                                                                                                                                           
\section{Introduction} 
\label{intro}

Spectacular outflows are often observed as a star forms. They occur during the phase when we expect inflows to  be dominant 
\citep[e.g.][]{2004orst.book.....S}. Many protostars have been discovered through their infrared jets while the driving internal activity is highly obscured  \citep{1994ApJ...436L.189M,1998Natur.394..862Z}. A widely accepted resolution is that the inflow is held up  by centrifugal forces in a rotating disc. 
The outflow then functions as the channel through which the angular momentum of the  disc
is gradually extricated. In support,
there are several plausible  mechanisms which may operate during accretion on to a single star 
to produce an outflow. These  lead to extended disc winds, magnetic tower flows and magnetospheric  X-winds   \citep{1982MNRAS.199..883B,1985PASJ...37...31S}. 
 
  We should not neglect that most stars
are not formed alone. For  each binary, we might expect up to three outflow systems associated with three  distinct discs with the large circumbinary disc feeding both a 
circumprimary disc and a circumsecondary disc. It follows that multiple outflow systems should be prevalent. Examples include the
jets from both components of the XZ\,Tau binary \citep{2008AJ....136.1980K}.
Multiple outflow systems   have also been  recently confirmed for L\,1551\.IRS5 \citep{Cruz_S_enz_de_Miera_2019}.  
Moreover, even the  older T\,Tauri binary V582\,Mon has been found to be surrounded by a circumbinary ring while driving a molecular outflow \citep{2018AJ....155...47A}. Yet, the observational evidence for multiple outflows remains very scattered. This leads us to ask how we should recognise multiple outflows.
The approach to the problem taken here is through numerical simulations. 

\begin{figure}
\centering
\includegraphics[width=1.0\linewidth]{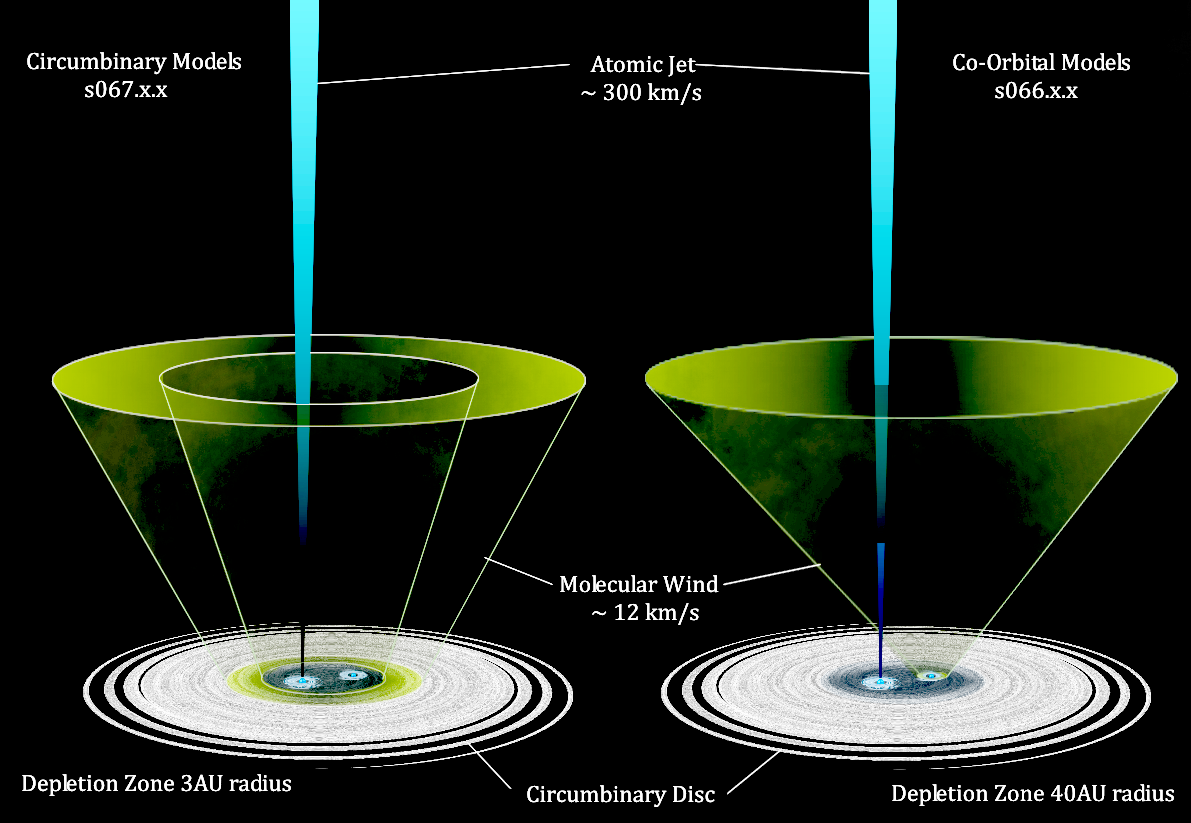}
\caption[Sketch of the Circumbinary and Co-Orbital Models]{An illustration  of the Circumbinary and  Co-orbital  Scenarios.  A simulation was performed within each model series  without the molecular flow as a control.}
\label{modelschema}
\end{figure}

We aim to identify the formation processes by analysing how the outflows release their kinetic energy during interactions.
The concept of orbiting outflows was first studied numerically by  \citet{2008A&A...478..453M}  where two adjacent atomic jets interact. Both jets  were narrow and
had a short dynamical flow time so that the long orbital time had little influence.
This was remedied in  Paper 1 \citep{2019MNRAS.tmp.2590L}   where the Co-orbital Scenario was fully explored. In that case, as illustrated in Fig.\,\ref{modelschema},  it is the two orbiting young stars  which are the effective origins.
The molecular outflow originates from from the secondary binary partner, as discussed by \citet{2008MNRAS.387.1313T}. The interaction is through direct impact  as a wide molecular wind clashes and spirals around the atomic jet. Regions of warm ionised gas are generated off-axis and this yield strong H$\alpha $ emission patches.
 
Here, we study the alternative  Circumbinary  Scenario in which the inner jet performs a short-period orbit,  remaining within a wide circumbinary outflow cone. This scenario was suggested by  \citet{2012AJ....144...61E}  as an interpretation for the HH\,30 outflows in which all the above components have been identified. 

The production of a collimated fast jet enclosed by a  wide low-speed outflow 
was uncovered in by  \citet{2008ApJ...676.1088M} in simulations. These outflows arise
during the very early protostellar stage associated with the outer adiabatic core and the second protostellar core.
This was confirmed   in three-dimensional magnetohydrodynamics simulation by 
\citet{2013ApJ...763....6T}, \citet{2014MNRAS.437...77B}, \citet{2014ApJ...796L..17M} and  \citet{2019ApJ...876..149M}.
 The jet was found to propagate within the outflow, forming a nested  structure in the velocity. The low-velocity outflow is in a nearly steady state, while the high-velocity jet appears intermittently due to a variety of inner disc instabilities.
  Although the jet carries the bulk of the  kinetic energy, the wide outflow channels the mass and momentum.  Here, we extend this work by studying the dual outflows and determining the physical properties which may lead to observable consequences.

In Paper 1 \citep{2019MNRAS.tmp.2590L}, we summarised the interpretations and parameters which have been previously found to
be consistent with both models as applied to HH\,30. We also detailed the  physical, chemical and  dynamical modifications  to the
ZEUS-MP code \citep{2006ApJS..165..188H}, and listed the full set of simulations. Radiation energy, magnetic flux and gravitational forces are neglected. The  full code includes a  three-species chemistry model for hydrogen that evolves H$_2$, H and H$^+$, according to the model presented in \citet{2007ApJS..169..239G}. The cooling of the gas includes that due to both atoms and molecules. The atomic cooling assumes a time-independent function and fine-structure cooling while molecular cooling includes vibrational and rotational contributions from H$_2$, CO, OH  and H$_2$O. Dissociative cooling and dust cooling are also included.

The above model concepts  are motivated by the   HH\,30 system which exhibits a well-collimated plume at visible wavelengths due to
hot atomic and partially ionised hydrogen gas  \citep{2018RMxAA..54..317G},  surrounded by a
colder, dense, wide-angle molecular  outflow \citep{2007ApJ...660..426H}.
In the Circumbinary  Scenario,   the binary system is configured to be very close and the wiggle in the atomic jet is caused by tidally-induced precession.  In this case, the molecular outflow is generated from the circumbinary disc, whose inner radius is  small (3\,AU) because of the tight binary orbit (0.75\,AU). 
  
  It should also be remarked that the orbital and disc rotation axes are taken to be aligned in this work. This probably minimises the interaction; misalignment between the outflow axes is likely  \citep{2019MNRAS.485.4667H} and merits further study. On the other hand, the remarkable collapse simulations described above have not quite reached the capability to yield  binary protostellar cores, discs and  sets of jets.
    
\section{Method}  
\label{methods}

Parameters for the circumbinary simulations were fixed by the following considerations.  Kinematic and morphological modelling of the   slow molecular outflow associated with HH\,30 by \citet{2008MNRAS.387.1313T} showed that the  
closest resemblance occurred when a very short-period binary orbit and a circumbinary disc source were taken. 
As a result, the jet wiggling is attributed to precession of the atomic jet source.
We adopt a set of parameters based on the HH\,30 outflow (defined as Type II in Table 4  of Paper~1).
 In summary, we take a tight binary with an orbital separation of 0.75\,AU. In this Type II circumbinary model,
 estimates  put the primary and secondary masses as
   $\sim$ 0.44$_\odot$ and $\sim$ 0.1 M$_\odot$   put the total mass at $\sim$ 0.54 M$_\odot$. The orbital period is one year. 
  
  The speed and mass flow rate of the injected  jet are  326 km\,s$^{-1}$ and 5.10\,$\times$\,10$^{-9}$\,M$_\odot$\,yr$^{-1}$.  The axis of this primary jet performs a motion in a cone as the driving disc is subject to precession with a period of 53 years with a half  angle of just 0.025 radians (1.43$^\circ$).
  
 The speed and mass flow rate of the injected  molecular wind are  11.7 km\,s$^{-1}$ and 6.15\,$\times$\,10$^{-8}$\,M$_\odot$\,yr$^{-1}$. 
  This wind, from a disc in fixed Keplerian rotation, stems from inner and outer radii of 9 -- 21\,AU.
  
    The speed of the jet is variable on short and long timescales. The long period variability in the speed of ejection would probably lead to the knots of bright emission observed  in the HH\,30 outflow on the scale of 0.1\,parsec   \citep{1990ApJ...364..601R}.  The short-period variability ($\sim$ months), which might be chaotic in nature, is triggered from variable accretion. The resulting compressions steepen into shock fronts that provide the collisional heating and ionisation of the jet material \citep{2007ApJ...660..426H,2007AJ....133.2799A}.
Furthermore, the    HH\,30 jet appears to be driven sideways on t he parsec scale in the sky plane. It displays a curved  West-facing 'C' shape  that may be caused by systemic intrinsic velocity changes, or impinging outflows or winds from nearby objects \citep{2012AJ....144...61E}.

The numerical calculations were performed using MPI with ZEUS-MP employing 180 tiles on the SCIAMA supercomputer. The desire for high resolution in three dimensions despite the strong lateral interaction restricted the grid length to just over 100\,AU. After much experimenting, we settled on the values provided in Table\,\ref{tab:modelgeometry}. In the x-dimension along the axis,  20 thin slabs each of 8 zones in width are stacked. Each of these slabs consist of a 3 x 3 arrangement of square 115 x 115 zone tiles in the y-z plane.

\begin{table*}
\resizebox{0.9\textwidth}{!}{\begin{minipage}{\textwidth}
  \centering
  \caption{Standard model geometry and tiling}
    \begin{tabular}{crrrrrcr}
    \hline
    Coordinate & Min   & Max   & Span  & Grid  & Zone Size & MPI   & Zones \\
          & (cm)  & (cm)  & (cm)  & Zones &  (cm) & Tiling & /Tile \\
    \hline
    x     & 0     & 1.600$\times 10^{15}$ & 1.600$\times 10^{15}$ & 160   & 1.0$\times 10^{13}$ & 20    & 8 \\
    y     & -1.725$\times 10^{15}$ & 1.725$\times 10^{15}$ & 3.450$\times 10^{15}$ & 345   & 1.0$\times 10^{13}$ & 3     & 115 \\
    z     & -1.725$\times 10^{15}$ & 1.725$\times 10^{15}$ & 3.450$\times 10^{15}$ & 345   & 1.0$\times 10^{13}$ & 3     & 115 \\
    \hline
    \end{tabular}%
  \label{tab:modelgeometry}%
\end{minipage} }
\end{table*}%

 Despite being inspired by HH\,30, the consequences of  the more general problem as sketched in the left panel of   Fig.\,\ref{modelschema} is of interest. To this end, a broad range of simulations were first performed until a small number of the most relevant conditions could be  selected for high resolution study.
Table \ref{tab:simulationruns} summarises the characteristics of the circumbinary simulations 
 designated by the prefix s067.
 Major long timescale (175 yrs) runs were performed to establish fully developed flow structures. The atomic jet model was also run  without the molecular flow.

 The  jet diameter is eleven cells, corresponding to  1.1\,$\times$ \,$^{14}$\,cm, commensurate with the assumed distance from the driving star of order 10\,AU. The initial temperature and ion fraction in the jet are fixed at arbitrary values with the pulses immediately raising them to appropriate vales. The circumbinary disc has an inner radius of  1.4\,$\times$ \,$10^{14}$\,cm, Mach number of 30 and a rotation speed of 6.5\,$\times$\,$10^{-9}$\,radians\,s$^{-1}$ which yields 9.1\,km\,s$^{-1}$. This creates a molecular wind with an opening angle  of $\sim$\,67$^\circ$.

 We then performed six additional runs of  shorter duration. These simulations; three with the velocity pulse time periods for the atomic jet varied and three with different values for the  orbital eccentricity taken.  The data retained for these simulations start from 65 years into the outflow evolution (the outflows having crossed the problem domain by this time and so display a fully developed flow) and traces the development for the ensuing 22 years.

\begin{table*}
  \centering
  \caption{Simulation Runs}
    \begin{tabular}{rrccrcrrrr}
    \hline
     &   {Simulations} & {Outflows} &     & Vpulse & Orbital &   &   {} \\
      &   & Atomic & Molecular & Period & $\epsilon$      & T$_1$ (yrs) & T$_2$ (yrs) \\
    \hline
          &       &       &       &       &       &       &       &       &  \\
          &       &       &       &       &       &       &       &       &  \\
     &{s067.3.2} & II    & II    & 5.256$\times 10^{6}$ & 0.00    & 0     & 175 \\
          &       &       &       &       &       &       &       &       &  \\
          & {s067.3.3} & II    & -     & 5.256$\times 10^{6}$& 0.00    & 0     & 175 \\
          &       &       &       &       &       &       &       &       &  \\
          & {s067.4.3} & -     & II    & & -       & 0     & 80 \\
          &       &       &       &       &       &       &       &       &  \\
          & {s067.4.3} & II    & II    & 1.051$\times 10^{7}$ & 0.00   & 65    & 87 \\
          & {s067.4.4} & II    & II    & 7.884$\times 10^{6}$ & 0.00    & 65    & 87 \\
          & {s067.4.5} & II    & II    & 1.314$\times 10^{7}$ & 0.00    & 65    & 87 \\
          &       &       &       &       &       &       &       &       &  \\
          & {s067.5.25} & II    & II    & 5.256$\times 10^{6}$ & 0.25   & 65    & 87 \\
          & {s067.5.50} & II    & II    & 5.256$\times 10^{6}$ & 0.50    & 65    & 87 \\
          &{s067.5.75} & II    & II    & 5.256$\times 10^{6}$ & 0.75   & 65    & 87 \\
          &       &       &       &       &       &       &       &       &  \\
    \hline
    \end{tabular}%
  \label{tab:simulationruns}%
\end{table*}%

The period of the pulsed velocity $T_{\rm vpulse}$ corresponds to the period of the  sinusoidal signal imposed on the velocity of injected material.  In all the simulations presented in these studies , the Relative Amplitude parameter $A_R$ used  is set to 0.2.  This generates a velocity signal of the form
\begin{equation}
V_J(t) = M \times C_J \times \frac{1+A_R\ \cos \omega_{v}t}{1+A_R},
\end{equation}

\noindent where $M$ is the jet Mach number, $C_J$ is the jet sound speed, and $\omega_{\rm v} = 2\pi/T_{\rm vpulse}$.  The maximum of this signal is $MC_J$ and the minimum is 66\% of this value.
The Keplerian orbit of the jet inlet is derived from the
orbital eccentricity parameter, $\epsilon$. 
The Newton-Raphson method is used in this calculation.
The dumps of HDF data  are saved at simulation time intervals of 2.125 $\times$ 10$^6$ seconds.  This value is chosen to be less than half of $T_{\rm vpulse}$.  This is the standard strategy
in signal processing:  in order to capture a sinusoidal signal the sampling frequency must be at least twice the highest frequency component of the signal.  In addition, with the choice of $T_S \approx 2.5 T_{\rm vpulse}$, we avoid `strobe' effects which can spoil animations. 
We may also proceed to determine proper motions of knots and other structures.

\section{Results}  
 
\subsection{A single atomic jet on a tight orbit}

\begin{figure}
\subfloat[$z-y$ plane, $x = 0$~cm]{\includegraphics[width=0.49\linewidth]{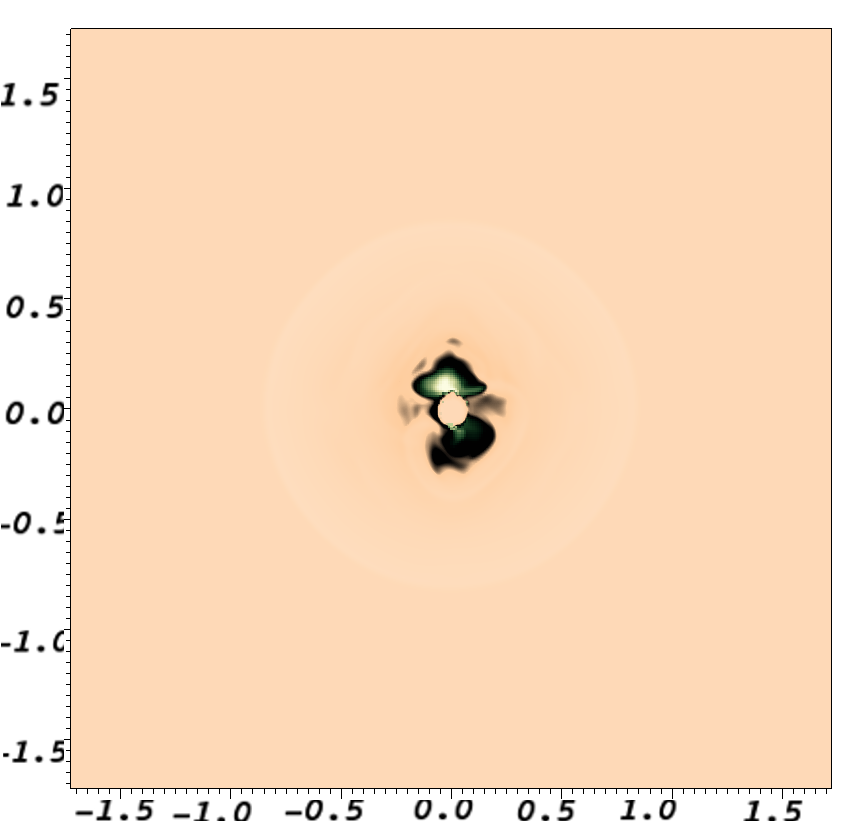}}
\subfloat[$z-y$ plane, $x = 7.5 \times 10^{14}~$cm]{\includegraphics[width=0.49\linewidth]{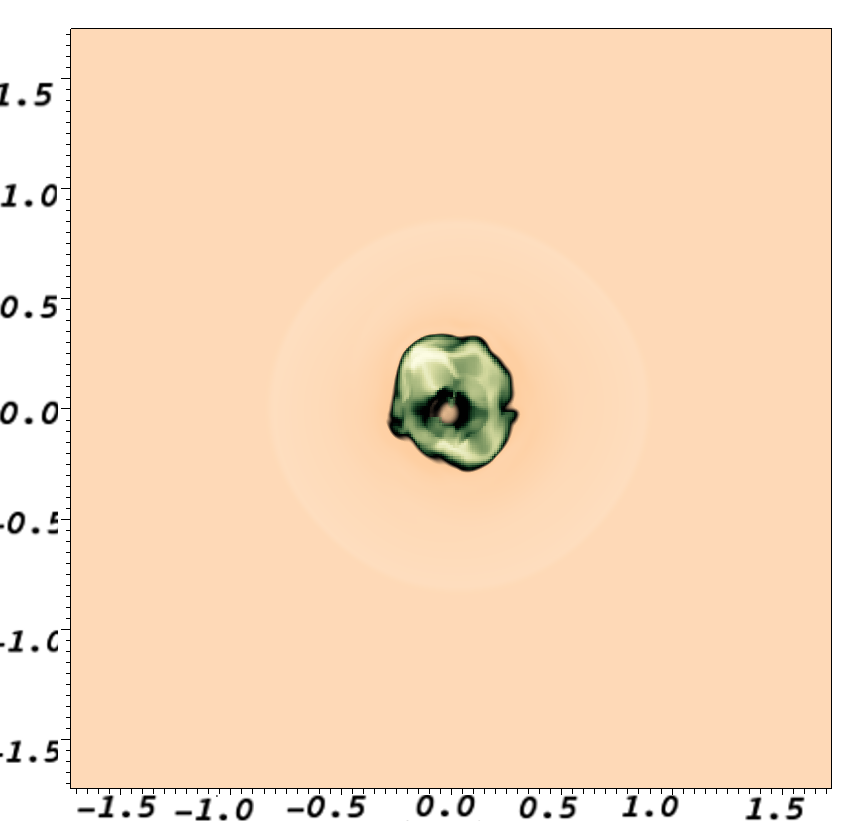}} \\\\
\subfloat[$z-y$ plane, $x = 1.5 \times 10^{15}$cm]{\includegraphics[width=0.49\linewidth]{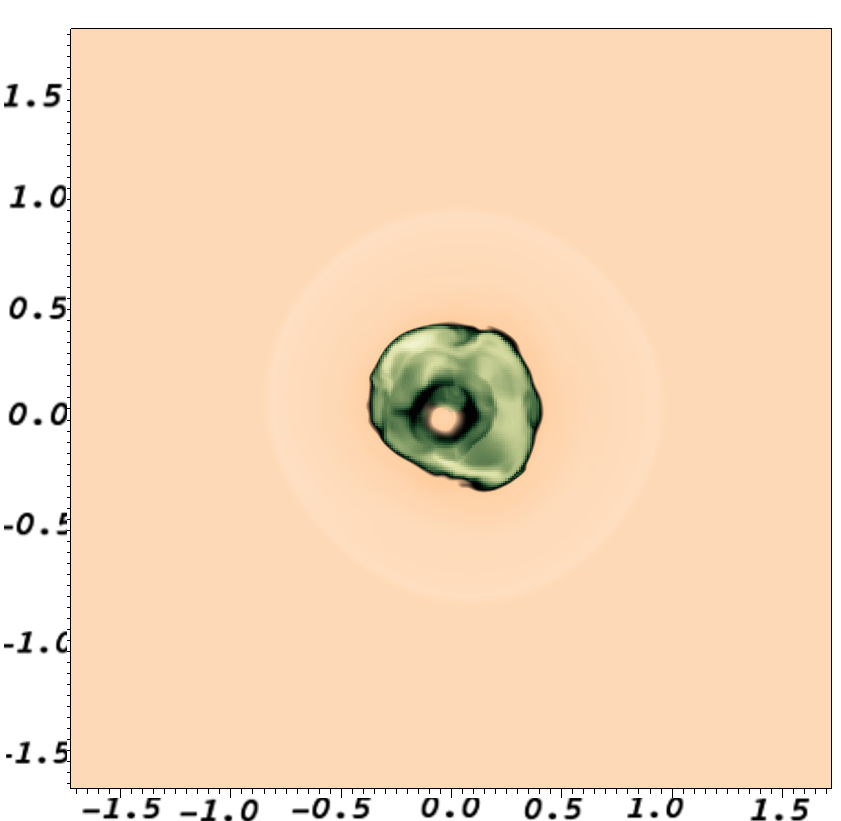}}\hspace*{0.02\linewidth} 
\subfloat[$x-y$ plane, z=0; $x-z$ plane, y = 0]{\includegraphics[width=0.49\linewidth]{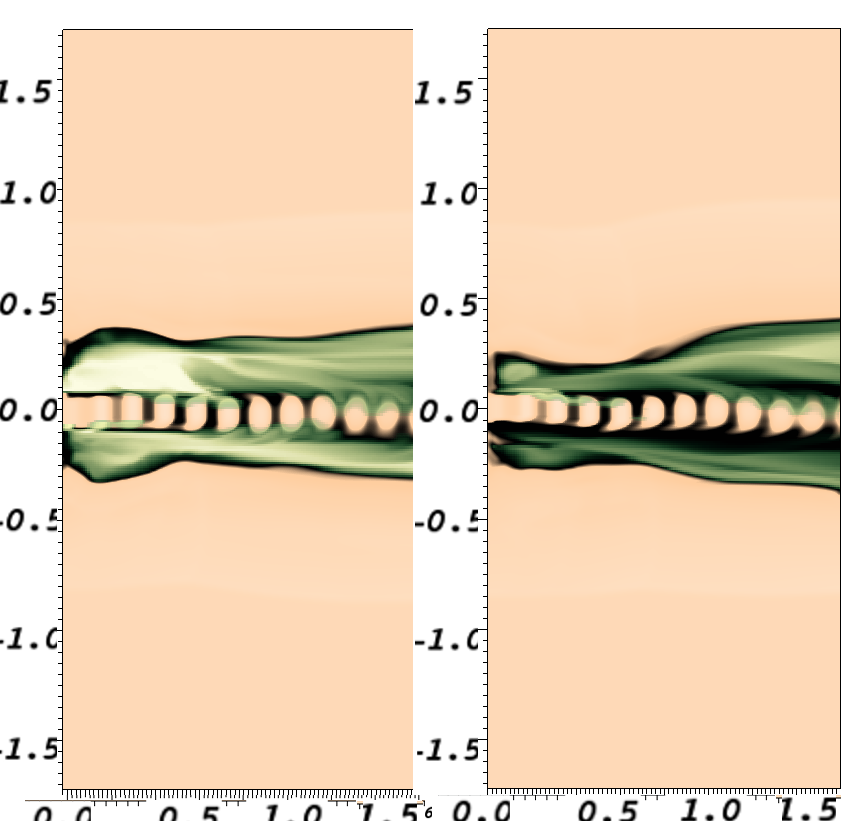}}\\\\
\hspace*{0.6cm} \subfloat[Density (g \, cm$^{-3}$)]{\includegraphics[width=0.30\linewidth]{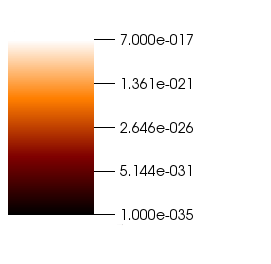}}
\subfloat[H$_2$ density (g \, cm$^{-3}$)]{\includegraphics[width=0.30\linewidth]{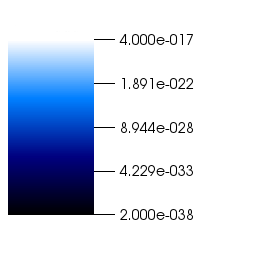}}
\subfloat[Fractional ion.]{\includegraphics[width=0.30\linewidth]{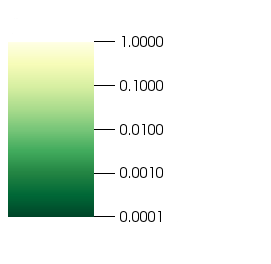}}
\\
\caption[Circumbinaryl Model s067.3.3: Section Plots, T=87.5 Years]{\textbf{Single atomic jet. } Cross-sections  under conditions of a  0.75\,AU binary,   at simulation time 87.5 years.  The axis scales are in units of 10$^{15}$cm.  The ambient medium is atomic  with traces of  molecular hydrogen formed  during the simulation.  The underlying density plot is fully opaque.  For clarity, H$_2$ (blue) and ionisation fraction (green) overplots are at ramped opacity: 100\% opacity at maximum value, fully transparent at minimum value.}
\label{6733_1300_xsect}
\end{figure}

The simulation designated as s067.3.3 models a scenario in which only the fast-moving, atomic jet is launched by the more massive binary partner in the two-star system.  The molecular outflow is absent. The immediately striking feature evident in the cross-sections of Figure\,\ref{6733_1300_xsect}  is the plume of partly  ionised gas filling the cocoon around the jet.
Shown in green on this figure, the ion fraction is about ten times higher than in the co-orbital equivalent. 

This broad ionised sheath has its explanation in the tight fast orbit which has the effect of spraying the jet over a wide area rather than allowing material to penetrate unhindered through an already evacuated narrow channel. The rapidly orbiting jet source spreads the ejected gas over a larger area. Remarkably, this leads to a wide conical distribution  for the ionised sheath  in the initial few AU, as shown in Panel (d), which could be mistakenly identified with the jet before collimation has been achieved. 

 The high excitation sheath was not found in Paper\,1.
 The difference between the present scenario and the co-orbital model  is the orbital period.  This model has a binary orbit separation of just 0.75\,AU, while the co-orbital jet had an 18\,AU separation.  Essentially, this jet remains in place and 'wiggles' furiously about the barycentre with a Keplerian orbital velocity of $\sim$ 26 km\,s$^{-1}$, while the long-period co-orbital jet moves gradually through the ambient medium at $\sim$ 5 km\,s$^{-1}$.
 
\begin{figure*}
\subfloat[density and composition at x=0\,cm]{\includegraphics[width=0.24\textwidth]{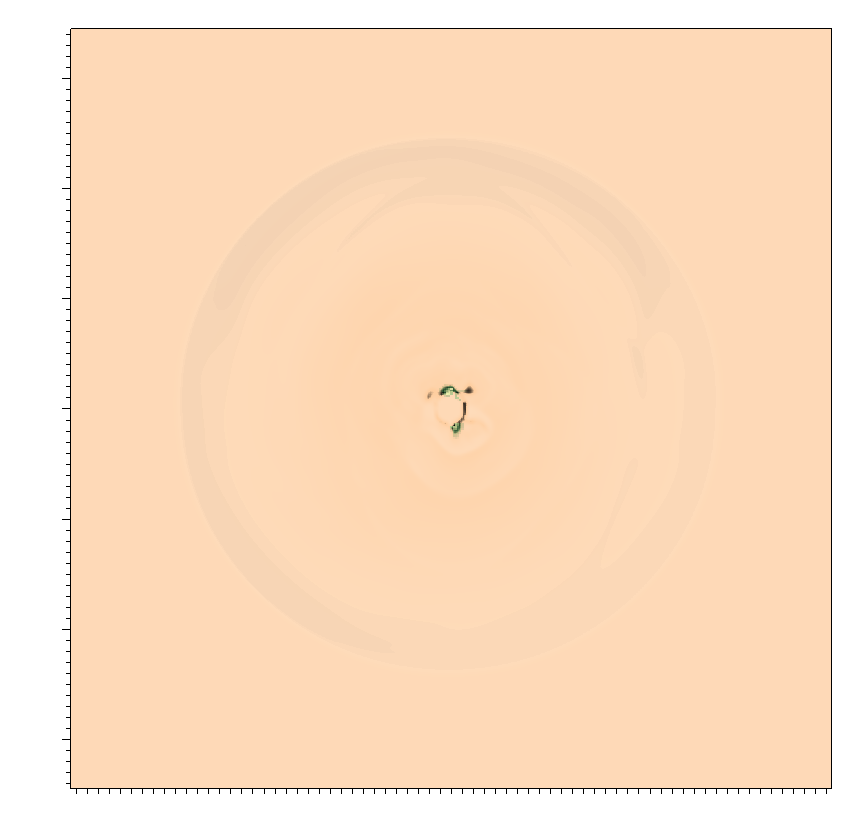}}
\subfloat[density and composition, 50AU]{\includegraphics[width=0.24\textwidth]{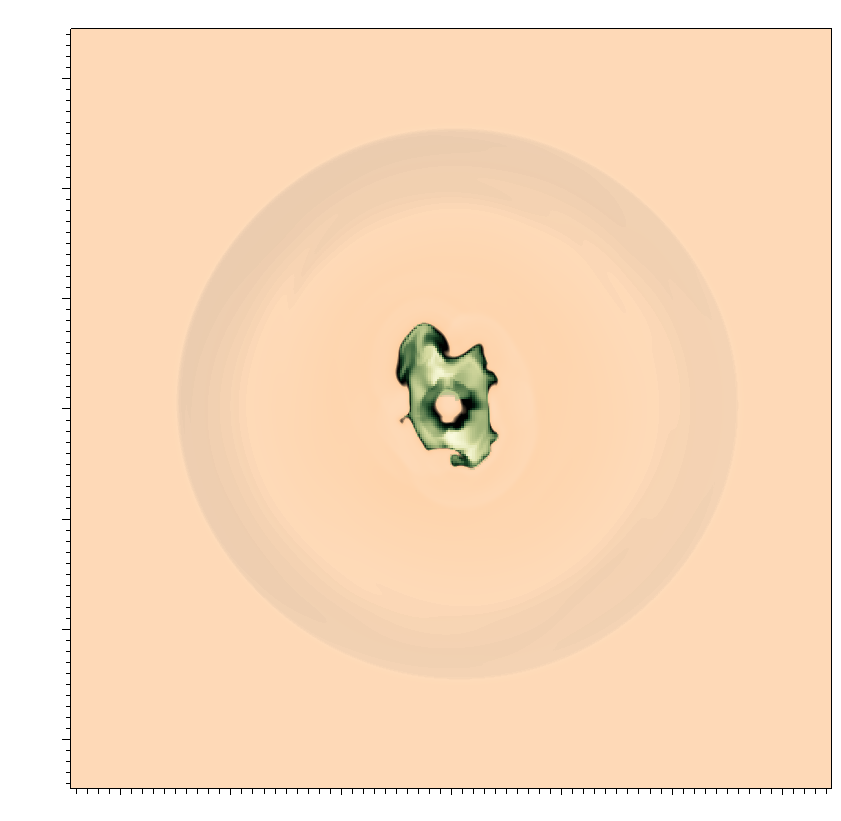}}
\subfloat[density and composition, 100AU]{\includegraphics[width=0.24\textwidth]{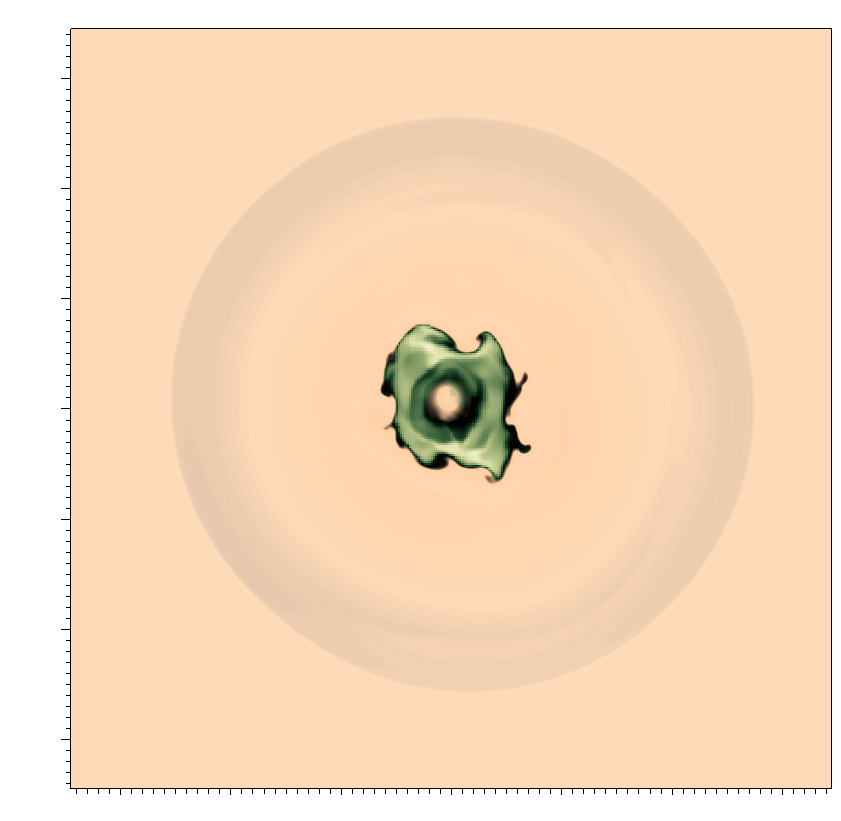}}
\subfloat[density and composition, xz plane]{\includegraphics[width=0.24\textwidth]{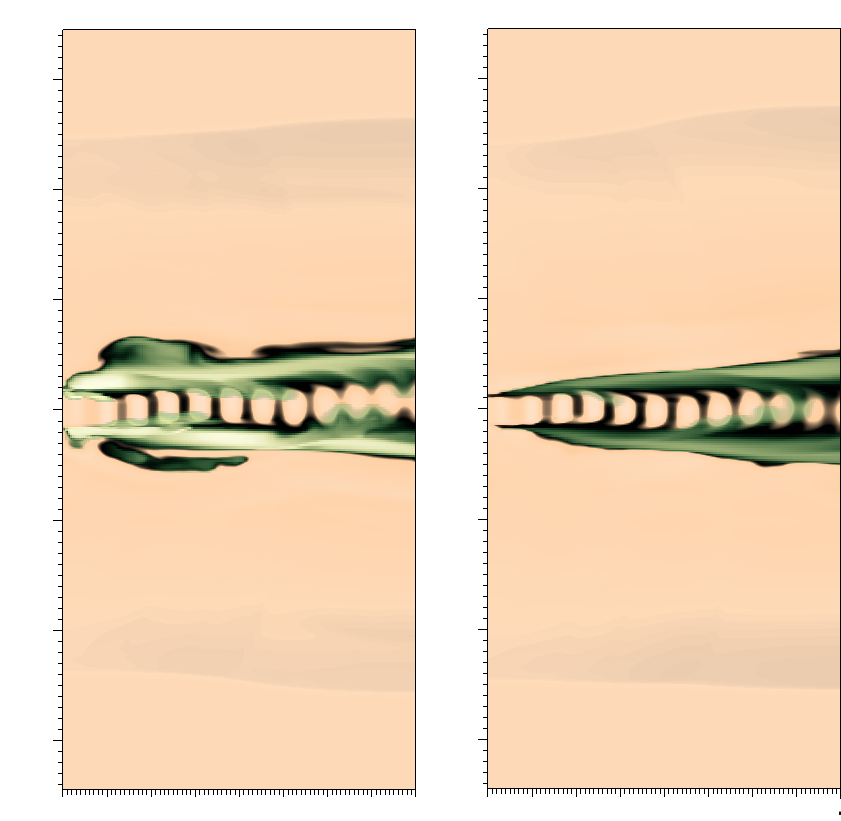}}\\
\subfloat[p, T and v at x=0\,cm]{\includegraphics[width=0.24\textwidth]{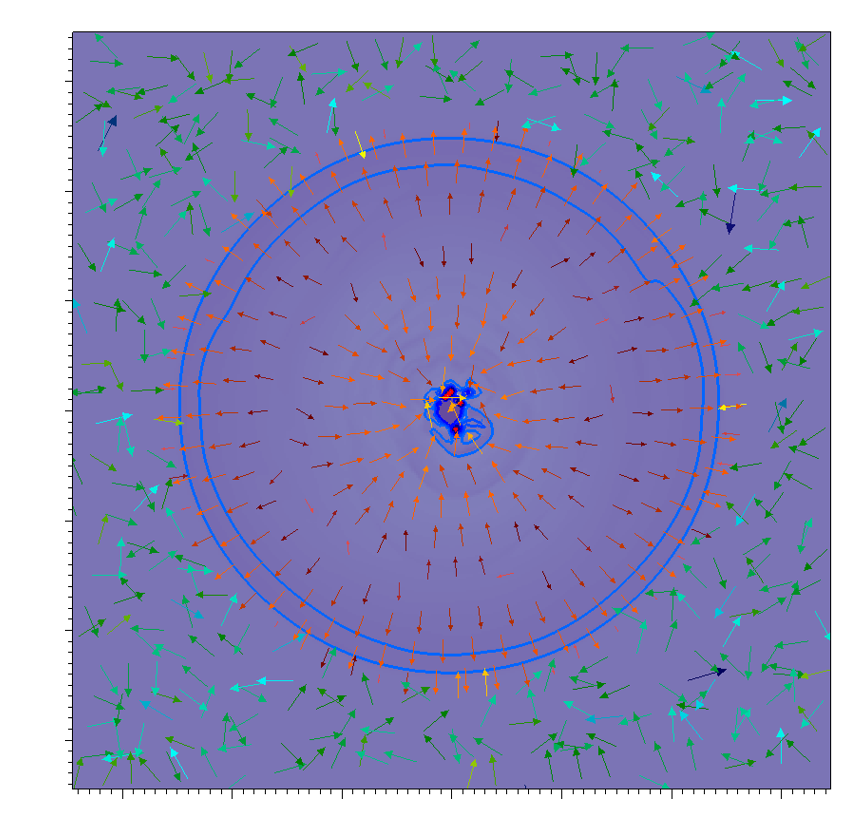}}
\subfloat[p, T and v at x=50AU]{\includegraphics[width=0.24\textwidth]{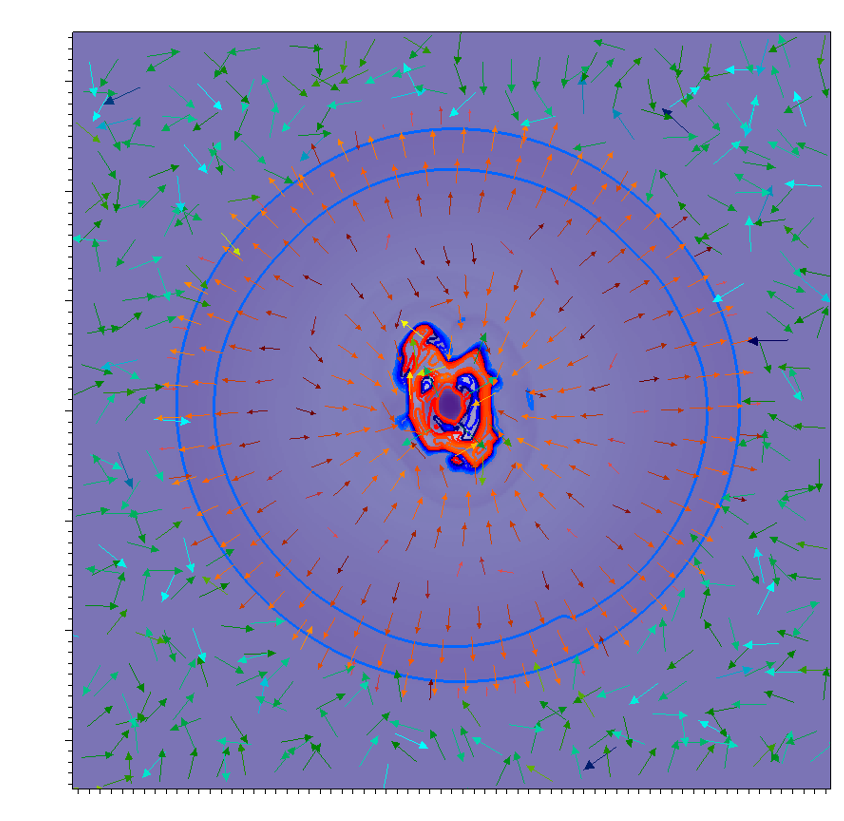}}
\subfloat[p, T and v at x=100AU]{\includegraphics[width=0.24\textwidth]{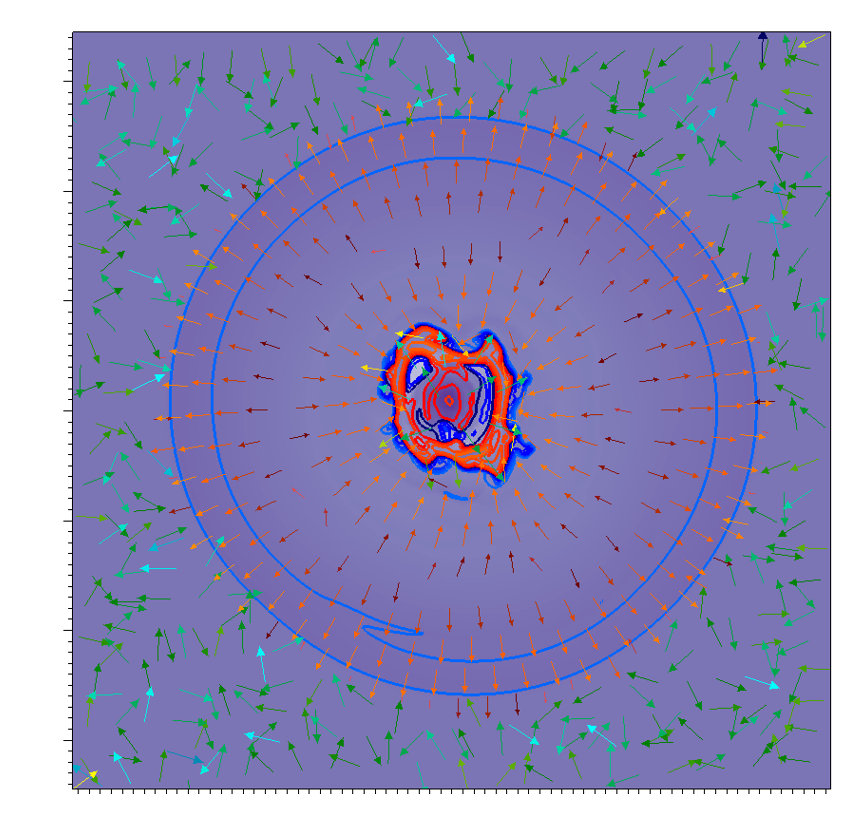}}
\subfloat[p, T and v in xy and xz planes]{\includegraphics[width=0.24\textwidth]{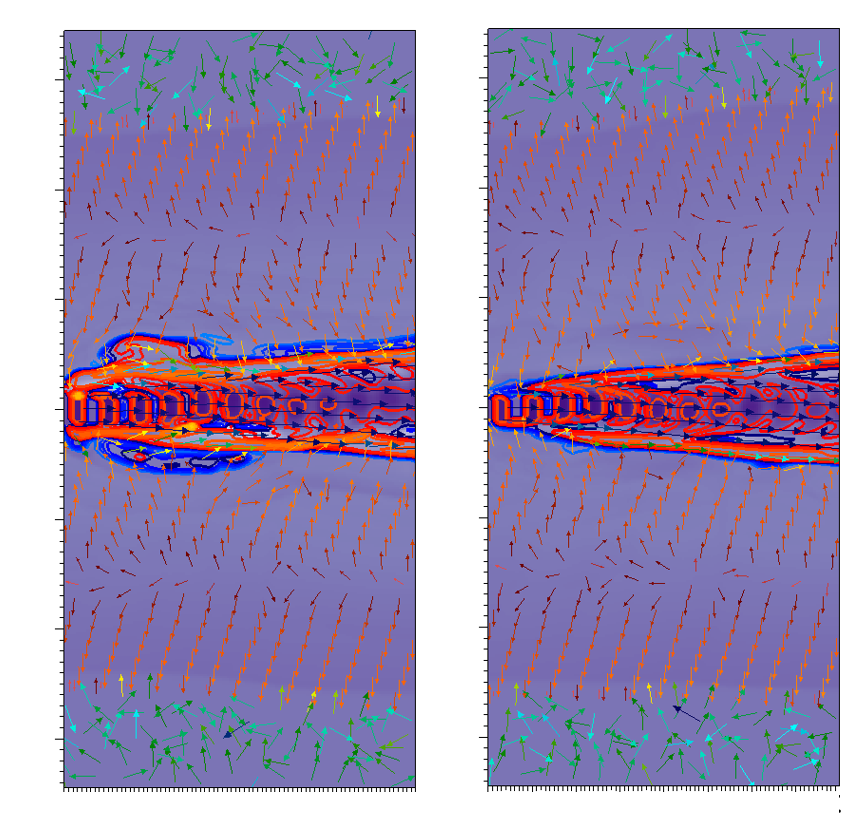}}
\\
\subfloat[density, H]{\includegraphics[width=0.16\linewidth]{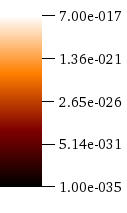}}
\subfloat[density, H$_2$]{\includegraphics[width=0.16\linewidth]{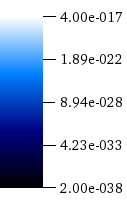}}
\subfloat[ion fraction]{\includegraphics[width=0.16\linewidth]{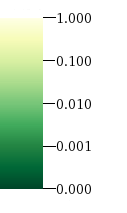}}
\subfloat[pressure]{\includegraphics[width=0.16\linewidth]{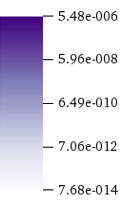}}
\subfloat[velocity vectors]{\includegraphics[width=0.16\linewidth]{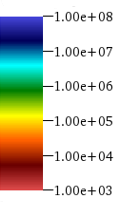}}
\subfloat[temperature contours]{\includegraphics[width=0.16\linewidth]{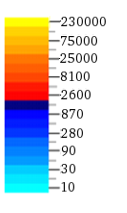}}\\
\caption[Co-Orbital Model s067.3.3: Section Plots, 175 years]{\textbf{Cross-sections of physical parameters for the single atomic jet}
 at the later simulation time of 175 years.     Axis scales are in units of 10$^{15}$cm.  Ambient medium is atomic hydrogen with trace molecular hydrogen formed during the simulation.  Underlying density plot is fully opaque. }
\label{single-pressure}
\end{figure*}

 A series of high-density knots rapidly develops within the jet  from the sinusoidal pulsed signal that begins purely as a velocity perturbation, as displayed also in Panels (d) and (h) of Fig.\,\ref{single-pressure} at the later fully-developed time.  There are eleven distinct density knots spanning the problem domain.  The ionisation fraction is low within the jet knots but there is a moderate increase towards the jet-cocoon interface.
 
 Pressure and temperature distributions   
  are also displayed in the lower panels of Fig.\,\ref{single-pressure}  for the the simulation time of 175 years.   
It is evident from these panels that the low-density cocoon surrounding the jet column, as well as the low-density regions within the jet, are populated by hot, partially ionised material. The interface with the ambient medium develops transverse structure probably related to the Kelvin-Helmholtz instability. The velocity along the jet direction
is suppressed within the sheath but the speed remains in  in the range 100\,--\,300\,km\,s$^{-1}$ across the jet.

\subsection{Dual Atomic-Molecular Outflow (simulation s067.3.2)}

\begin{figure}
\subfloat[z-y plane, $x = 0$\,cm]{\includegraphics[width=0.49\linewidth]{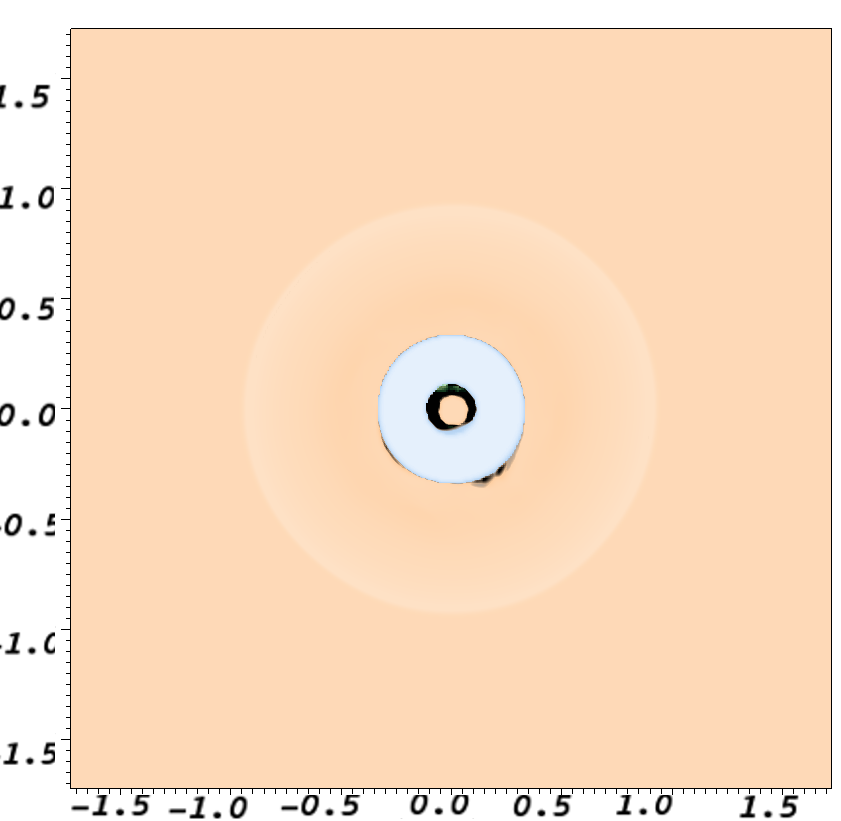}} 
\subfloat[z-y plane, $x = 7.5 \times 10^{14}$ cm]{\includegraphics[width=0.49\linewidth]{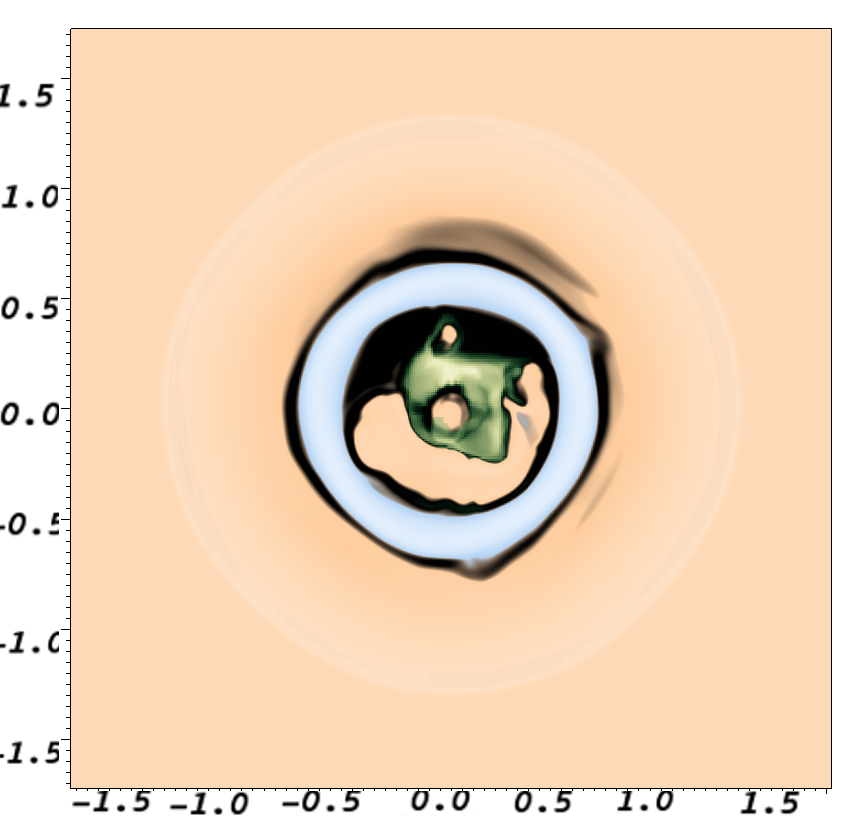}} \\\\
\subfloat[z-y plane, $x = 1.5 \times 10^{15}$ cm]{\includegraphics[width=0.49\linewidth]{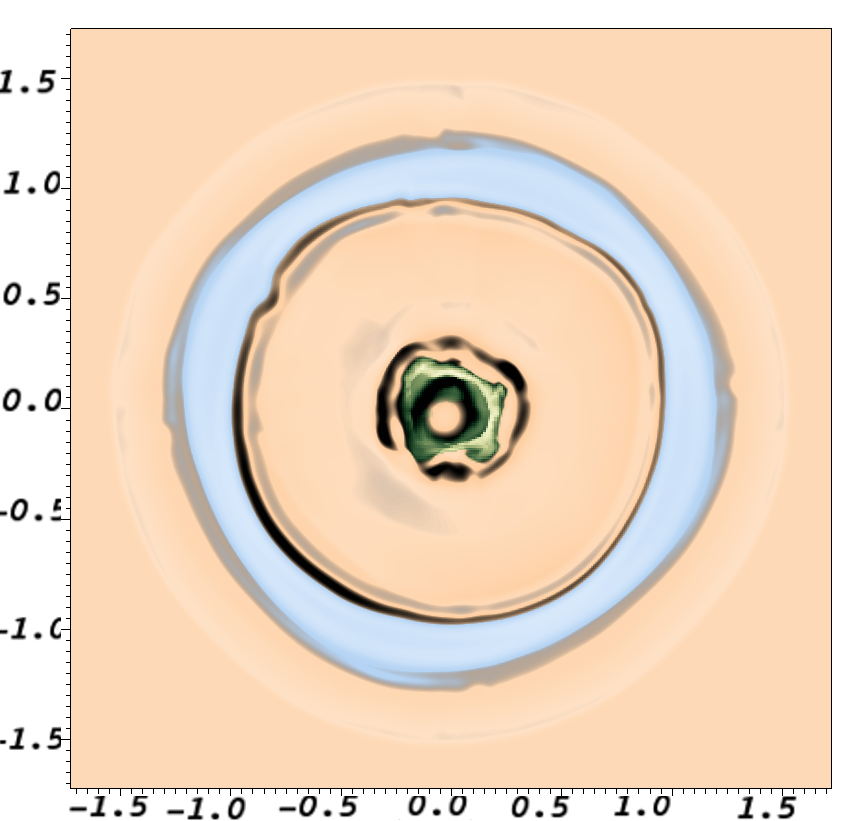}}
\subfloat[x-y plane, z = 0cm; $x-z$ plane, $y = 0$cm]{\includegraphics[width=0.49\linewidth]{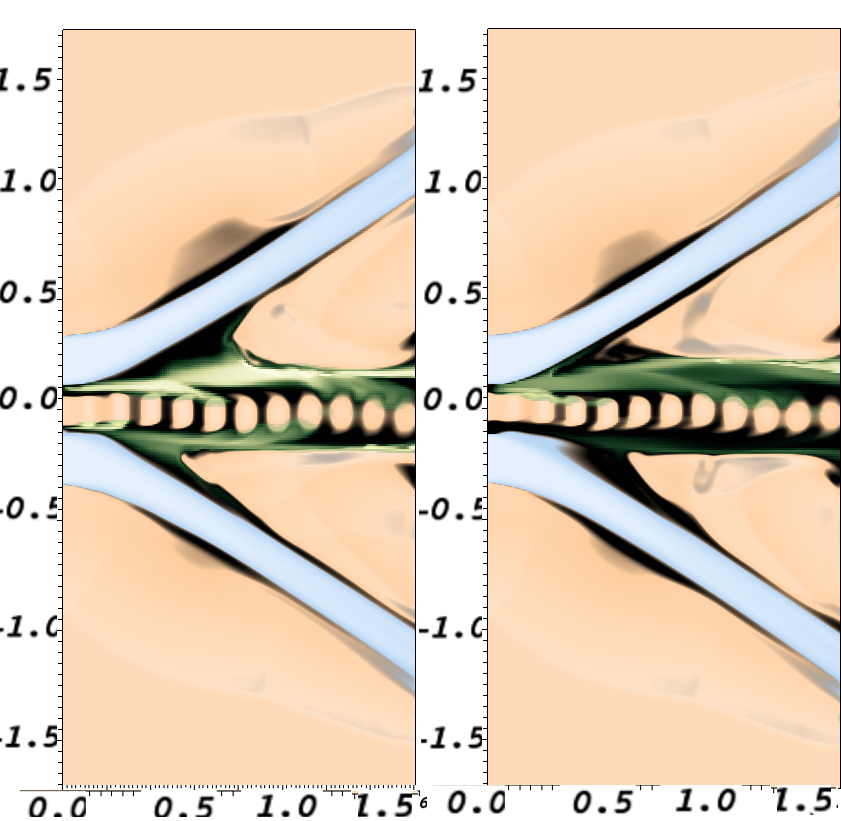}}\\\\
\hspace*{0.6cm} \subfloat[log(density, gcm$^{-3})$]{\includegraphics[width=0.30\linewidth]{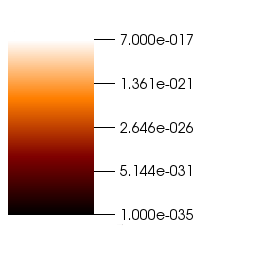}}
\subfloat[log(H$_2$ density, gcm$^{-3}$)]{\includegraphics[width=0.30\linewidth]{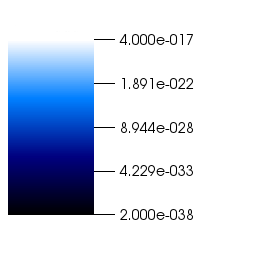}}
\subfloat[log(ion fraction)]{\includegraphics[width=0.30\linewidth]{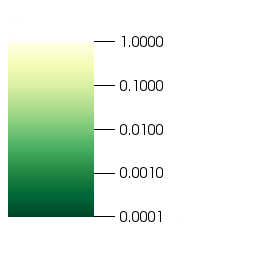}}
\\
\caption[CCircumbinary Model s067.3.2 Section Plots, 87.5 years]{\textbf{Circumbinary Model s067.3.2: 0.75\,AU Binary, atomic-molecular outflow, density  cross-sections at simulation time 87.5~years.}  Axis scales are in units of 10$^{15}$cm.   Underlying density plot is fully opaque.  For clarity, H$_2$ and ionisation fraction overplots are at ramped opacity; 100\% opacity at max value, transparent at minimum value.}
\label{6732_1300_xsect}
\end{figure}

The molecular outflow, emerging from the circumbinary disc, is now introduced along with the above atomic jet.
 The set of  cross-sections of density  is displayed in Figure \ref{6732_1300_xsect}  for the simulation time  of 87.5 years. All the figure share a common colour scaling of variables which permits a direct comparison.

The morphology of the molecular outflow strongly supports the ad hoc and ballistic particle models calculated  by  \citet{2006A&A...458..841P} and  \citet{2008MNRAS.387.1313T}. The lateral expansion is driven mainly by the Keplerian rotational velocity profile rather than the pressure differential, although the outflow is 1.5 $\times$ overpressured with respect to the ambient medium. This an important contrast with the Co-orbital dual-outflow model.  We see the expanding remnant of the bow shock in the ambient medium as before but this is now  dominated by the broader bow shock of the molecular outflow.

 The features of the inner part of the central pulsed atomic jet column are virtually identical to those already identified in the atomic-only case.  The ionised sheath, however, is compressed from the base by the proximity to the high-pressure disc outflow. This has squeezed the sheath which then remains relatively narrow, as shown in Panel (d) of Figure \ref{6732_1300_xsect}. 
 
\begin{figure*}
\subfloat[density and composition at x=0\,cm]{\includegraphics[width=0.24\textwidth]{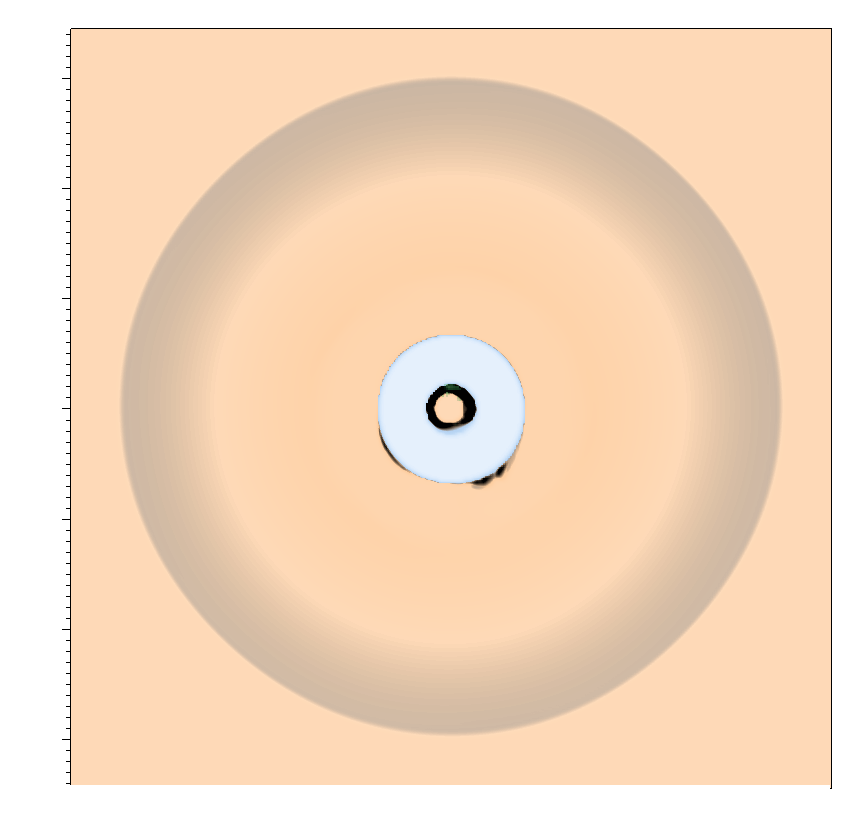}}
\subfloat[density and composition, 50AU]{\includegraphics[width=0.24\textwidth]{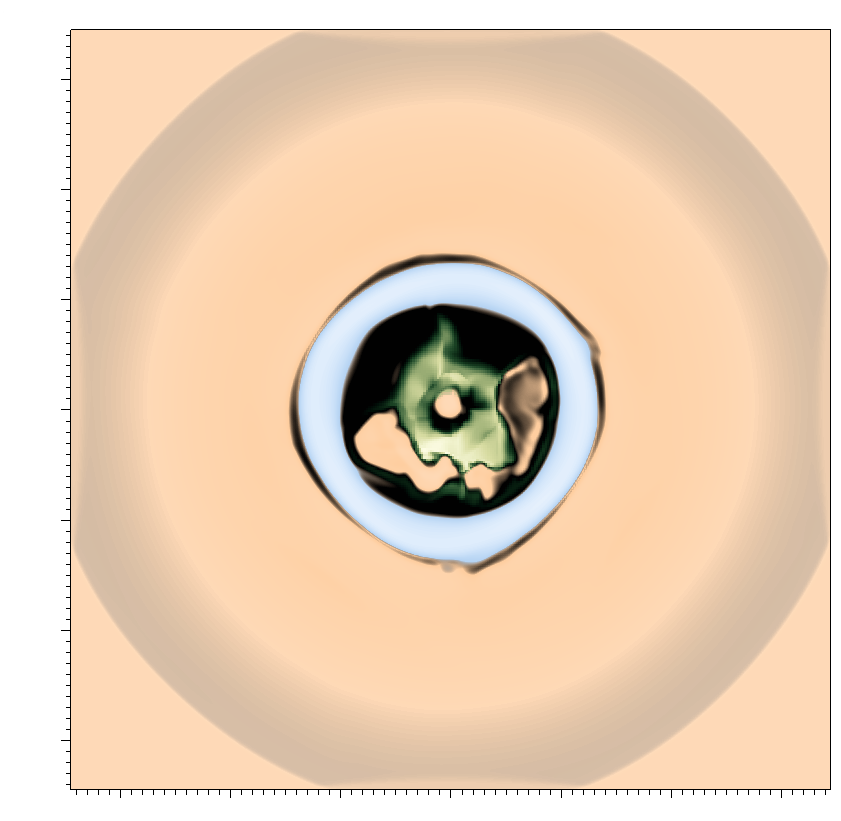}}
\subfloat[density and composition, 100AU]{\includegraphics[width=0.24\textwidth]{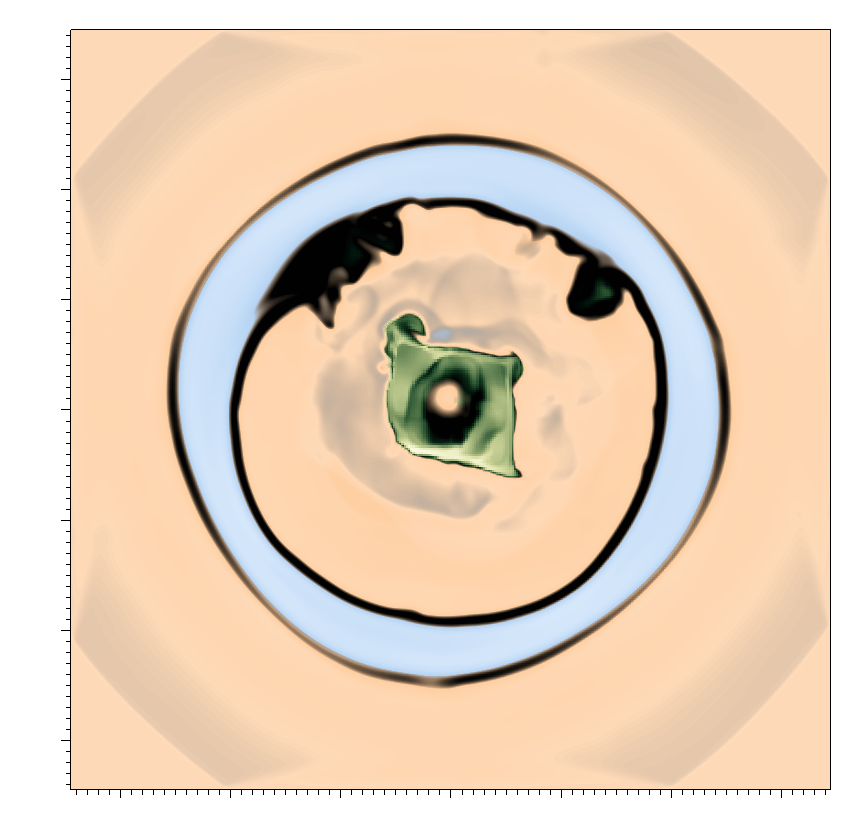}}
\subfloat[density and composition, xz plane]{\includegraphics[width=0.24\textwidth]{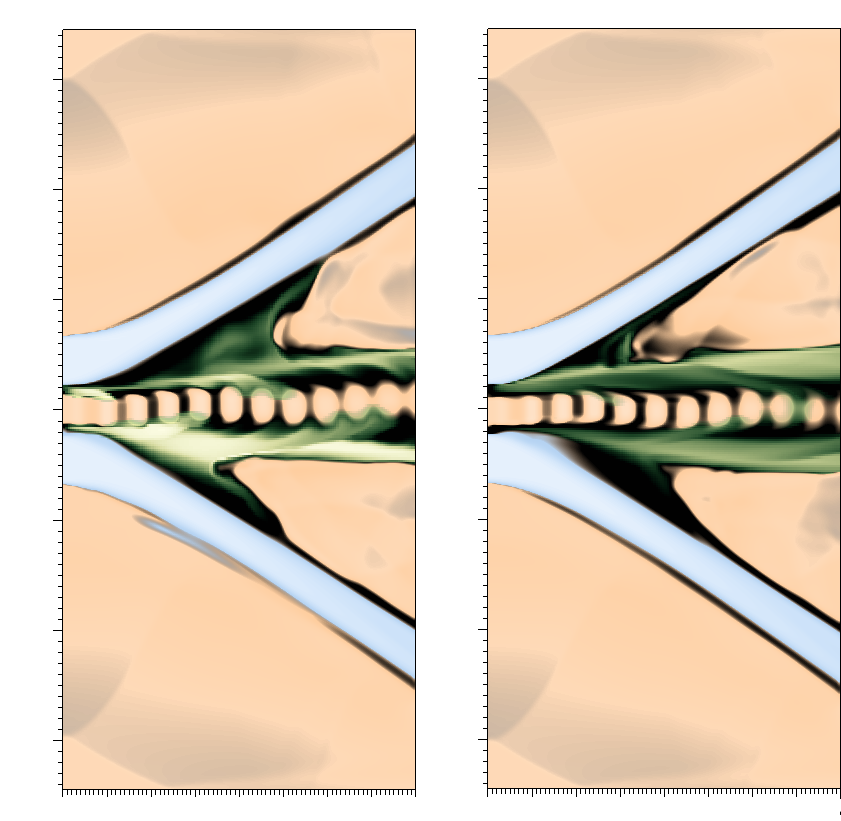}}\\
\subfloat[p, T and v at x=0\,cm]{\includegraphics[width=0.24\textwidth]{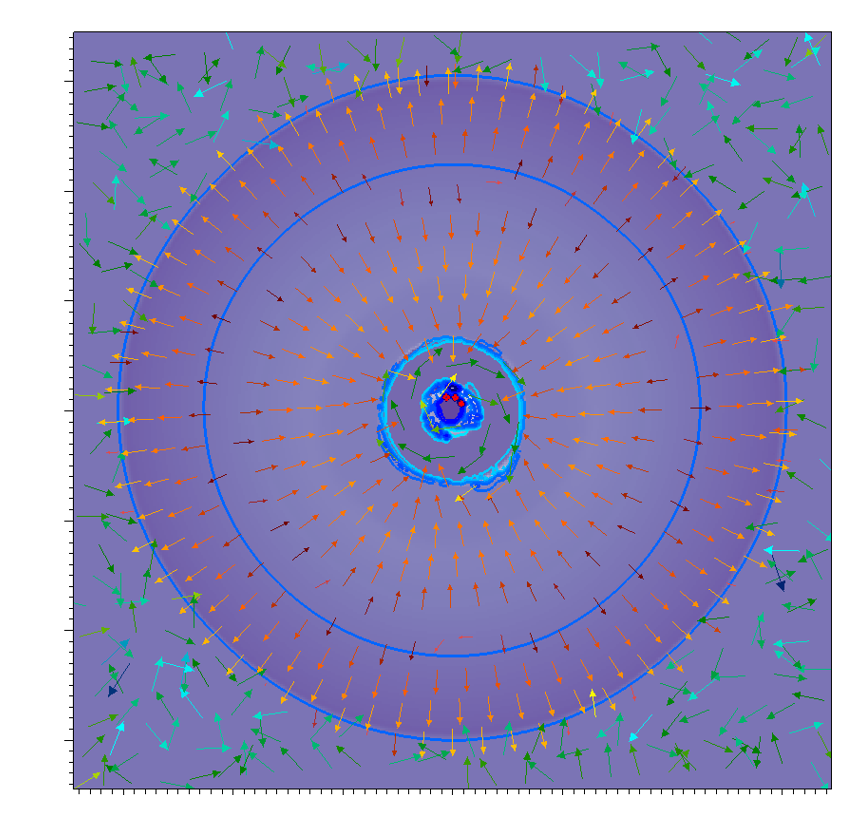}}
\subfloat[p, T and v at x=50AU]{\includegraphics[width=0.24\textwidth]{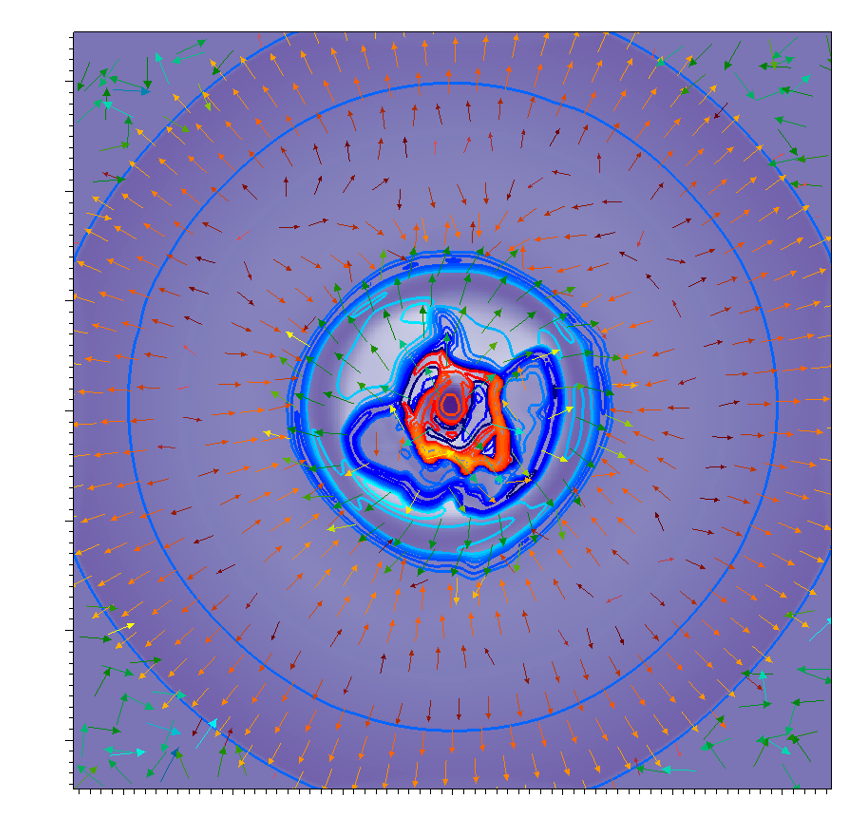}}
\subfloat[p, T and v at x=100AU]{\includegraphics[width=0.24\textwidth]{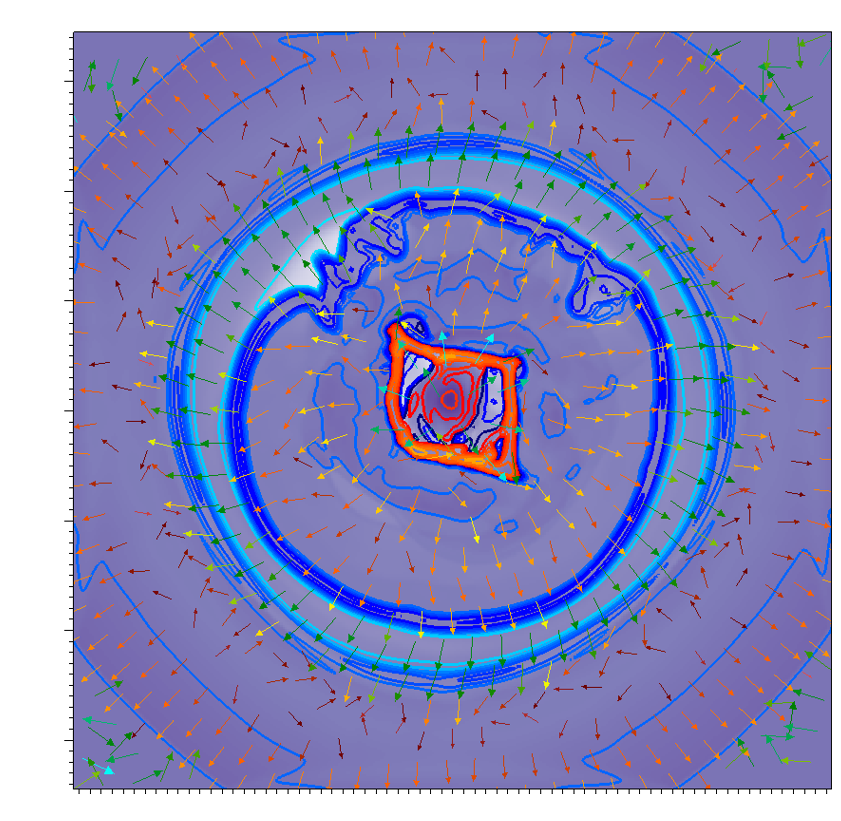}}
\subfloat[p, T and v in xy and xz planes]{\includegraphics[width=0.24\textwidth]{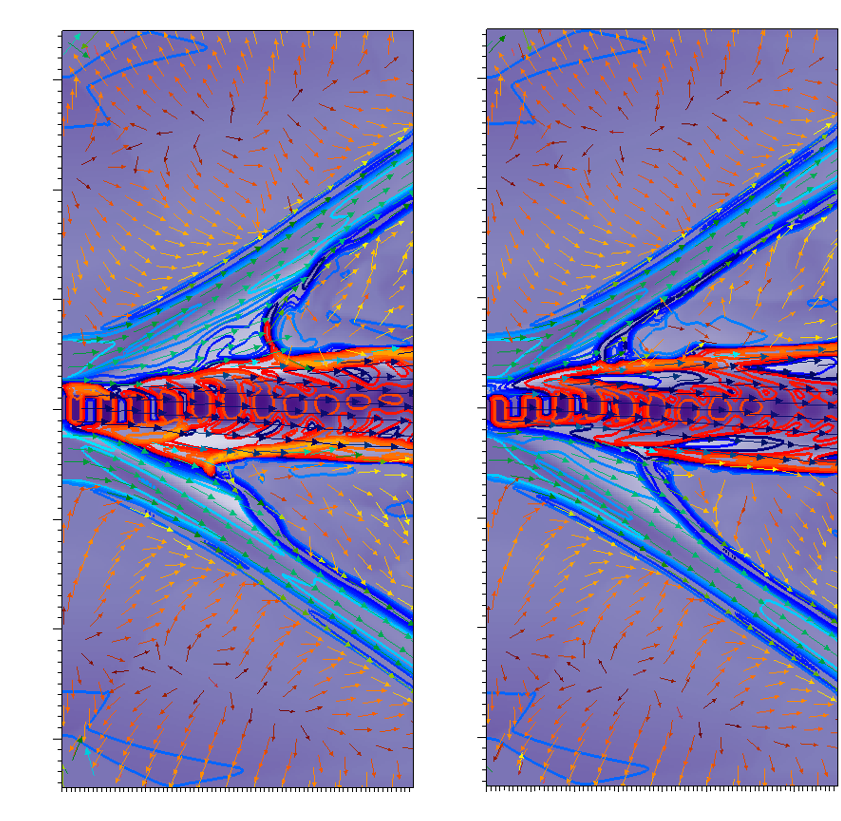}}
\\
\subfloat[density, H]{\includegraphics[width=0.16\linewidth]{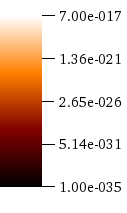}}
\subfloat[density, H$_2$]{\includegraphics[width=0.16\linewidth]{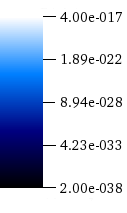}}
\subfloat[ion fraction]{\includegraphics[width=0.16\linewidth]{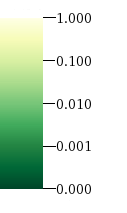}}
\subfloat[pressure]{\includegraphics[width=0.16\linewidth]{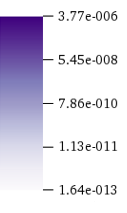}}
\subfloat[velocity vectors]{\includegraphics[width=0.16\linewidth]{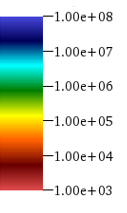}}
\subfloat[temperature contours]{\includegraphics[width=0.16\linewidth]{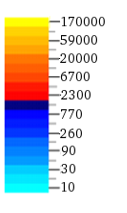}}\\
\caption[Model  Section Plots, 175 years]{\textbf{Cross-sections of physical parameters for the Circumbinary Scenario} atomic-molecular outflow at simulation time 175 years.  Axis scales are in units of 10$^{15}$cm.  Ambient medium is atomic hydrogen with trace molecular hydrogen formed during the simulation.  Underlying density plot is fully opaque. }
\label{6732_2599}
\end{figure*}

 In Panel  (b) we see that a low density cavity has been carved out from the ambient medium enclosed by the conical molecular flow.  This kind of evacuated region does not have an equivalent in the atomic-only  simulation.  The ambient medium is trapped, prevented from flowing back into the cavity by waves of hot, light, partially ionised  material expanding out from the atomic jet periphery and impacting against the boundary of the denser material.  As the simulation progresses (see Fig.\,\ref{6732_2599}) the cavity is slowly inflated as the ambient material is eroded.

\subsection{Temperature and Excitation}

\begin{figure*}
\centering
\includegraphics[width=0.85\linewidth]{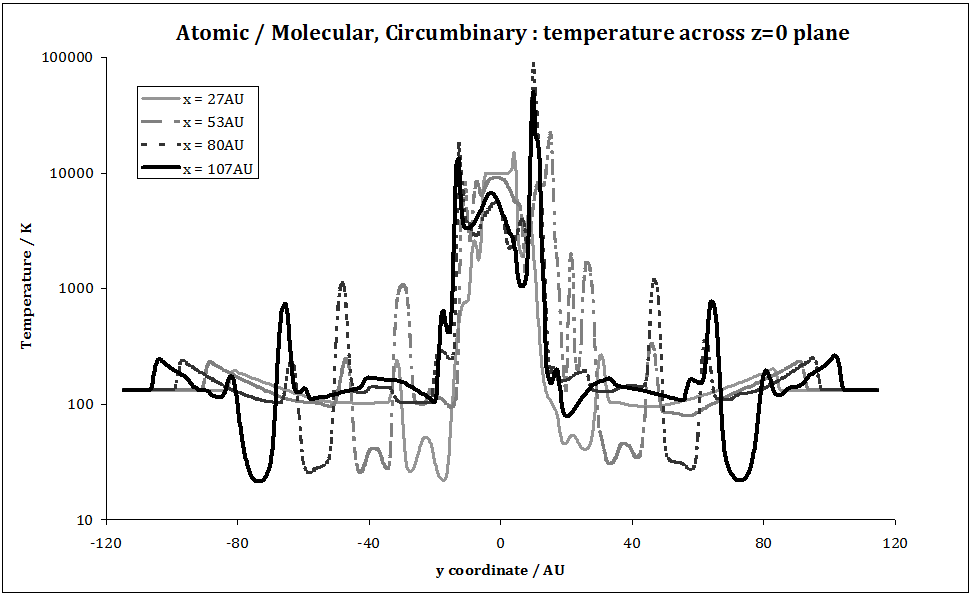}
\caption[Circumbinary atomic/molecular model, temperature profile]{Temperature profiles across the z=0 central plane at selected distances from the inlet (x-axis values) for the  circumbinary model s067.3.2.}
\label{temperatureprofiles}
\end{figure*}

 The temperature profiles across the lateral span of the domain are shown in Figure \ref{temperatureprofiles} at three distances along the barycentric (x) axis.  Note the logarithmic temperature scale.  The outer 'shoulders' correspond to the bow shock from the molecular outflow, propagating outwards at the medium sound speed.  
 
Deep troughs in the temperature profile correspond to the cold molecular material from the circumbinary disc while the central high plateau shows the presence of the jet.  Note that the region immediately surrounding the jet is hotter than the jet itself.  The very high temperatures developed in the periphery of the jet increase with distance from the jet inlet boundary (see legend), while the temperature in the jet column is decreasing. Clearly, the warm atomic jet maintains a constant width of $\sim$ 20\,AU while the narrow shear layer in which kinetic energy has been dissipated separates the atomic jet from the ambient cavity. 

\begin{figure*}
\centering
\includegraphics[width=0.85\linewidth]{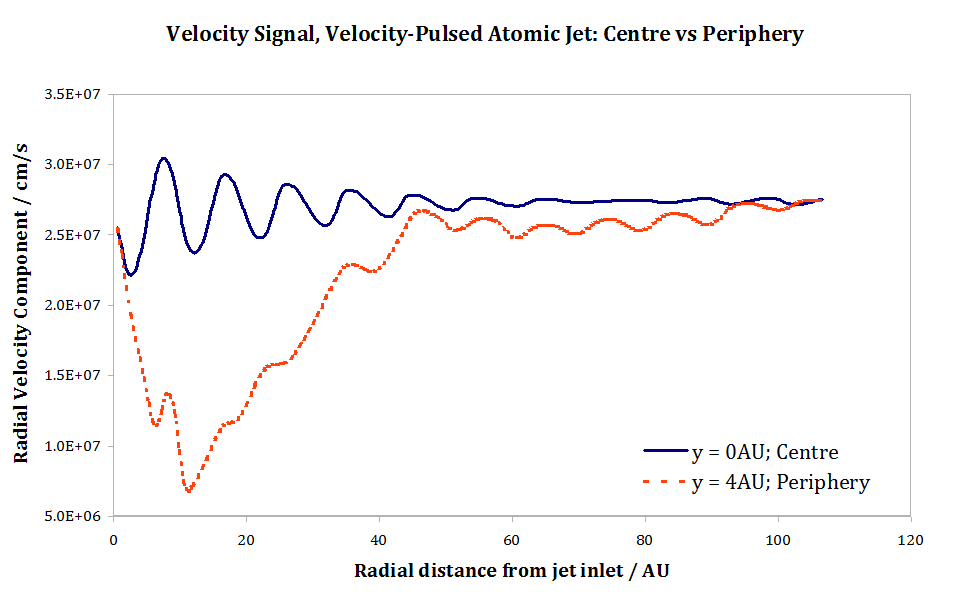}
\caption[Centre-line vs Peripheral velocity signals: pulsed atomic jets]{Centre-line vs Peripheral velocity signal along the axis of the pulsed atomic jet in the dual atomic-molecular, circumbinary model s067.3.2.    These can be seen to correspond to the emerging ionisation region in the bottom left corner of Fig.\ref{6732peripheryion}}
\label{vsignalcentre}
\end{figure*}

In order to understand the ionisation process in this model, it is instructive to examine the velocity behaviour of material in the jet as it orbits tightly about the barycentre.  Figure \ref{vsignalcentre} compares the centre-line velocity signal down the middle of the jet with the velocity variation on the jet periphery.

The velocity signal at the jet inlet is 2.717 $\pm$ 0.543 $\times$ 10$^7$ cm/s varying sinusoidally.  The centre-line velocity eventually settles at a mean value of 2.746 $\pm$ 0.014 $\times$ 10$^7$ cm/s at the outgoing boundary; though there is some suggestion in the signal that the velocity amplitude might be re-intensifying in line with the findings of \citet{1997A&A...320..325S}.  It is evident that the velocity difference at the jet periphery is greater than within the jet column itself with an initial steep fall of 54\% from the value at the inlet boundary over the first 6\,AU followed by a small recovery before another sharp fall to 73\% below inlet value by 11AU.

\begin{figure*}
\begin{minipage}[t]{0.9\textwidth}
    \vspace{0pt}
    \centering
\subfloat{\includegraphics[width=0.97\linewidth]{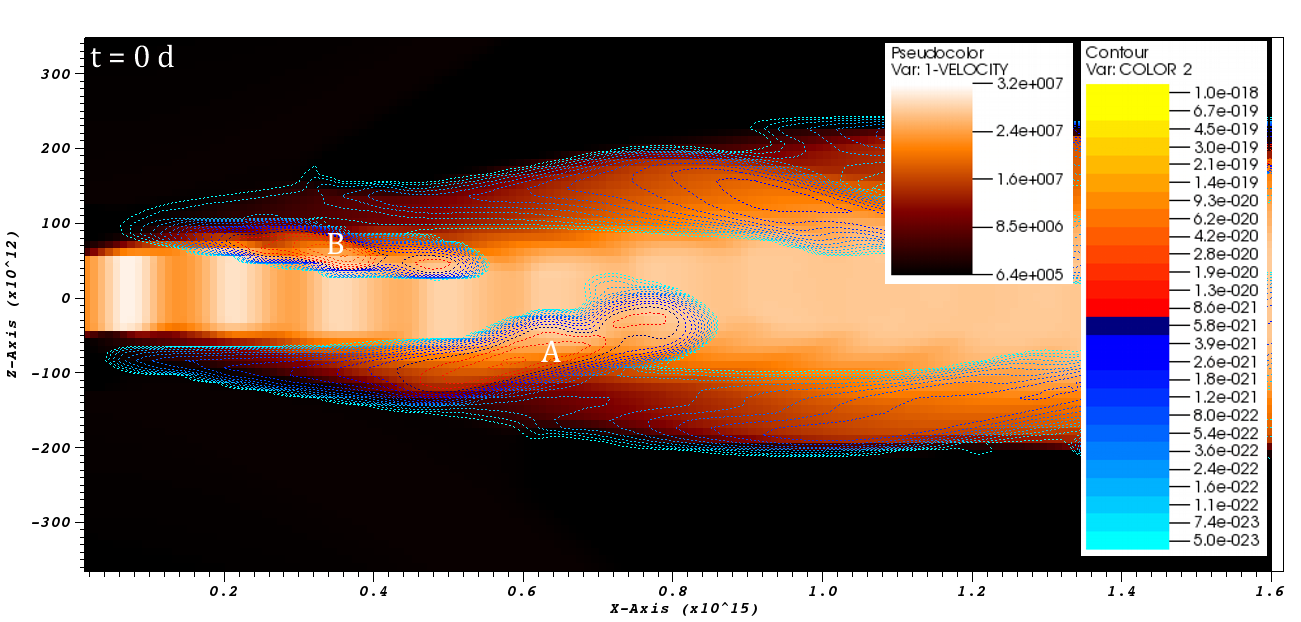}}\\[-1ex]
\subfloat{\includegraphics[width=0.97\linewidth]{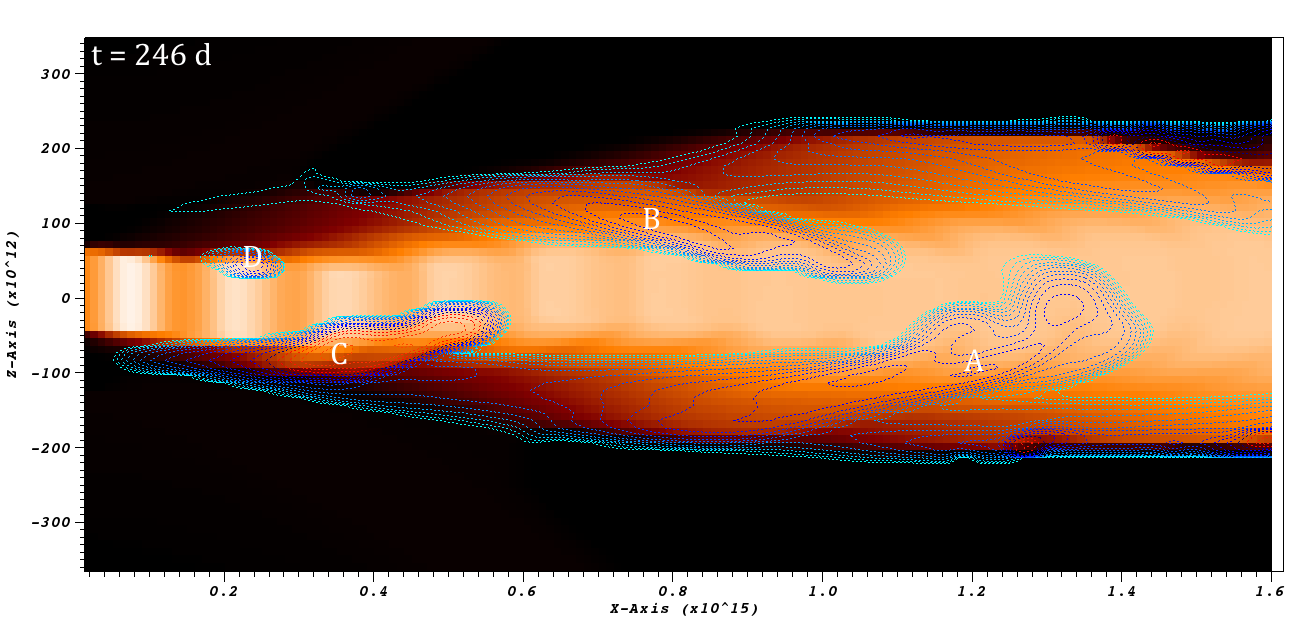}}\\[-1ex] 
\subfloat{\includegraphics[width=0.97\linewidth]{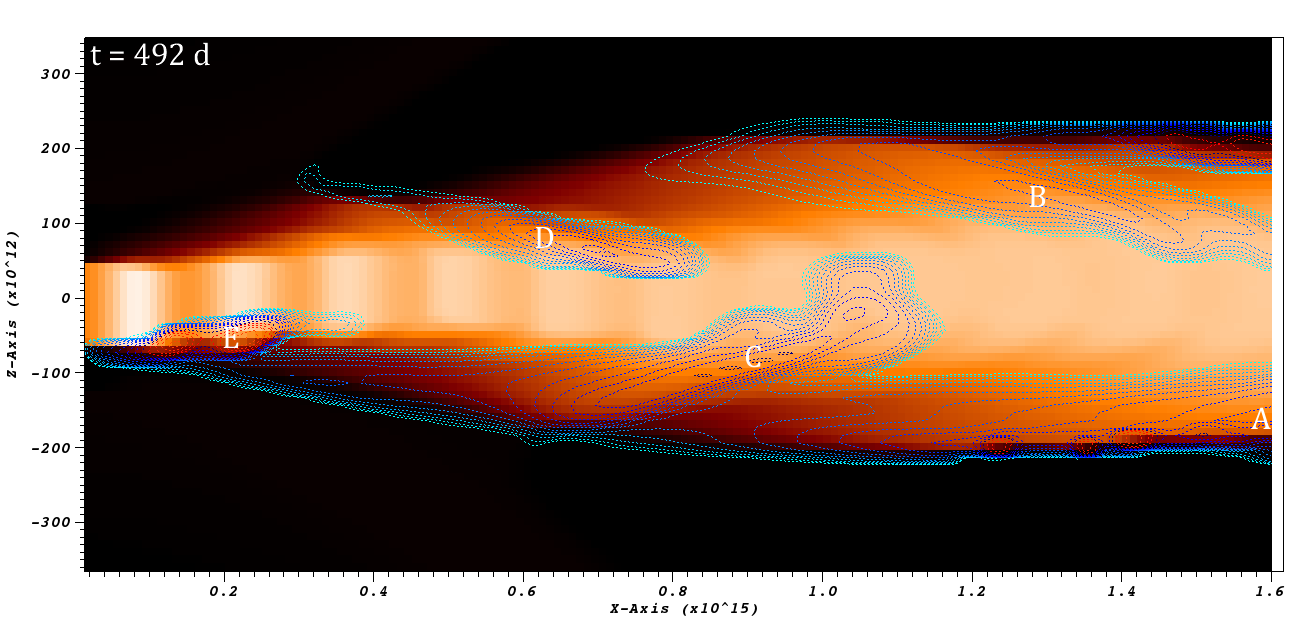}}
  \end{minipage}\hfill
\caption[Circumbinary Model: formation and propagation of ionised regions]{Formation and propagation of ionised regions on the periphery of the atomic jet in the model s067.3.2. Fluid elements  A, B, C, D are regions of strong ionisation; in fact cross-sections through a continuous shock that winds around the jet column.   Underlying colour plot is the V$_x$ component (cm\,s$^{-1}$). 
Overlaid contours (g\,cm$^{-3}$) correspond to partial density of ionised Hydrogen. Cross-sections are at z = 0 and axes     are cm scale. }
\label{6732peripheryion}  
\end{figure*}

This explains how the ionised sheath comes to envelop the jet column in the Circumbinary Scenario (including s067.3.3 which has no molecular component).  As the centre of the 3.67\,AU radius jet inlet describes a 0.75\,AU orbit about the barycentre, the periphery of the jet is subject to a continual strong oblique shock, almost as abrupt as the main jet bow shock, in contact with the ambient medium that refocuses the jet inwards.   The expansion of the hot gas drives a crossing shock inwards towards the jet centre.  Furthermore this shock is a feature that screws around the jet as the motion of the jet carries the disturbance downstream. Figure \ref{6732peripheryion} shows this explicitly for individual fluid elements that we have tracked as they are convected downstream.

\begin{figure*}
  \begin{minipage}[t]{0.14\textwidth}
    \vspace{0pt}
\subfloat{\includegraphics[width=1.0\linewidth]{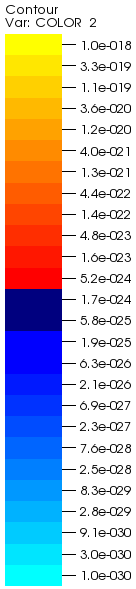}}\\[-1.5ex]
\subfloat{\includegraphics[width=1.0\linewidth]{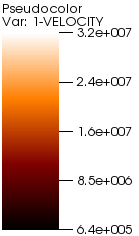}}
  \end{minipage}\hfill
  \begin{minipage}[t]{0.85\textwidth}
\vspace*{3cm}
    \vspace{0pt}
\subfloat{\includegraphics[width=0.98\linewidth]{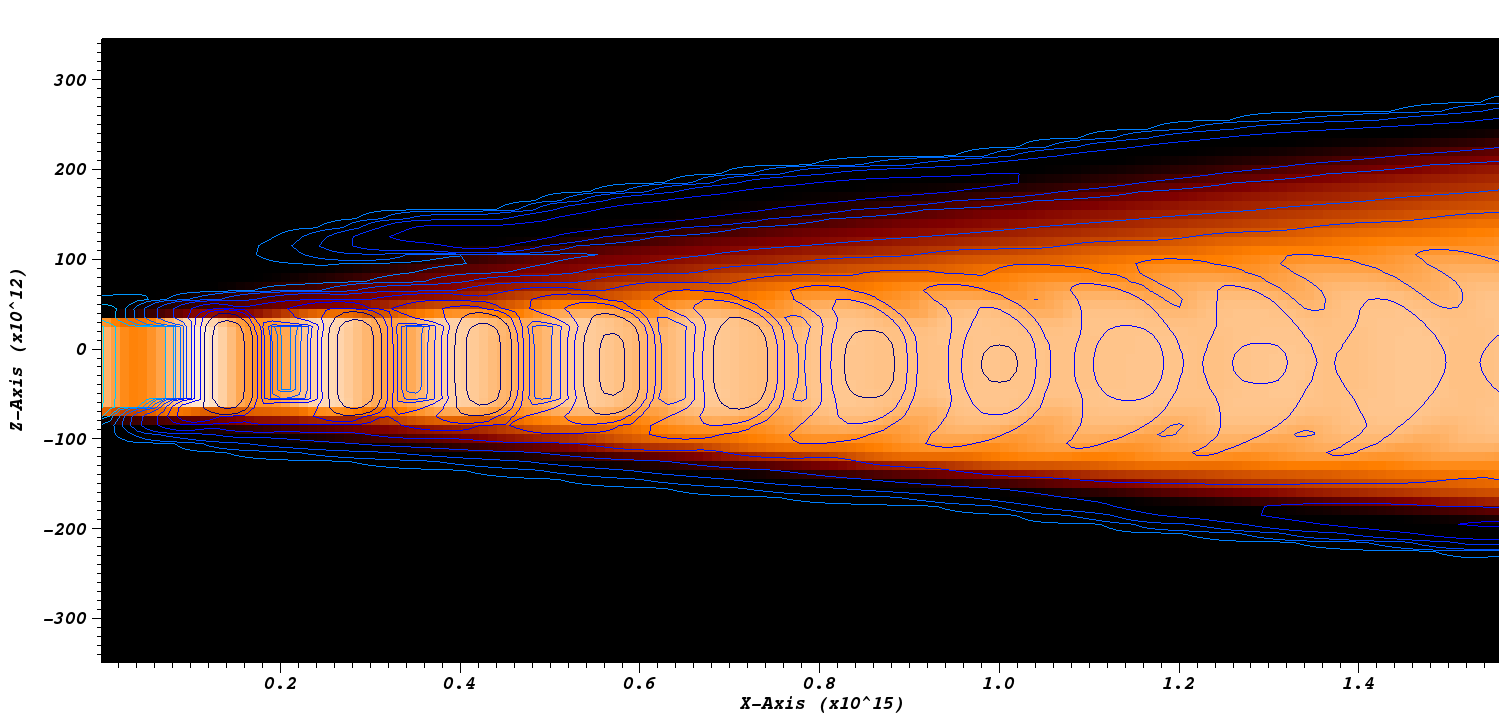}}\\[-6ex]
\subfloat{\includegraphics[width=0.98\linewidth]{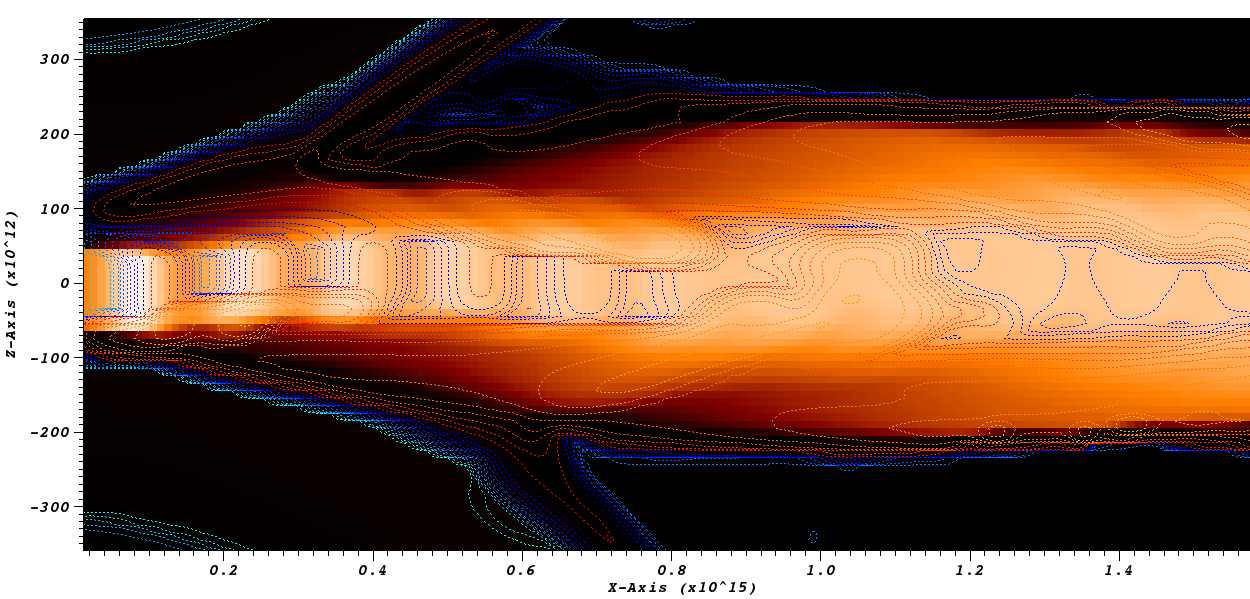}}
  \end{minipage}
\vspace{-0.5em}\caption[Comparison of ionisation modes, co-orbital vs circumbinary]{Comparison of ionisation modes in Co-orbital Scenario s066.2.4 (upper panel) vs Circumbinary Scenario s067.3.2 (lower panel).  Colour plot is the V$_x$ component (cm s$^{-1}$); contours (g cm$^{-3}$) are partial density of ionised Hydrogen; axes are cm scale.  In Fig.\ref{6732peripheryion}, the contour scale excludes regions of low ionisation to highlight the peripheral screw-thread ionisation shock; here the range is expanded to include lower ion densities.  The weakly ionised 'arms' branching off from above and below the jet in the circumbinary model are material entrained into the boundary layer of the slow-moving molecular flow.}
\label{6766ion}
\end{figure*}

Can this model be distinguished from the Co-orbital Scenario? A direct comparison is displayed in Fig\,\ref{6766ion}  where the ion density
is displayed as contours. There are three major differences. Firstly, the circumbinary model generates a jet which is less streamlined in axial velocity although after 100\,AU the difference in width is not an identifying difference. Secondly, the ion density is much higher within the sheath of the atomic jet, typically by a factor of 100. This generates observable emission as will be elucidated below. Thirdly,
there are weakly ionised arms branching off above and below the jet in the circumbinary model which represents material entrained into the boundary layer of the slow-moving molecular flow. This is mainly due to the proximity of the disc which has an inner edge at 9\,AU from the binary barycentre.
 
\subsection{Longitudinal Analysis of Circumbinary Models}
\label{longmethod}

The jet and molecular outflow do not strongly interact in the Circumbinary Scenario. This contrasts with the co-orbital case of Paper\,1 where the jet has to penetrate through the molecular cone as it sweeps around. The result is that the analysis of 
quantities  in one-dimensional plots along or transverse to the axis is relatively simple.

\begin{figure}
\centering
\includegraphics[width=1.0\linewidth]{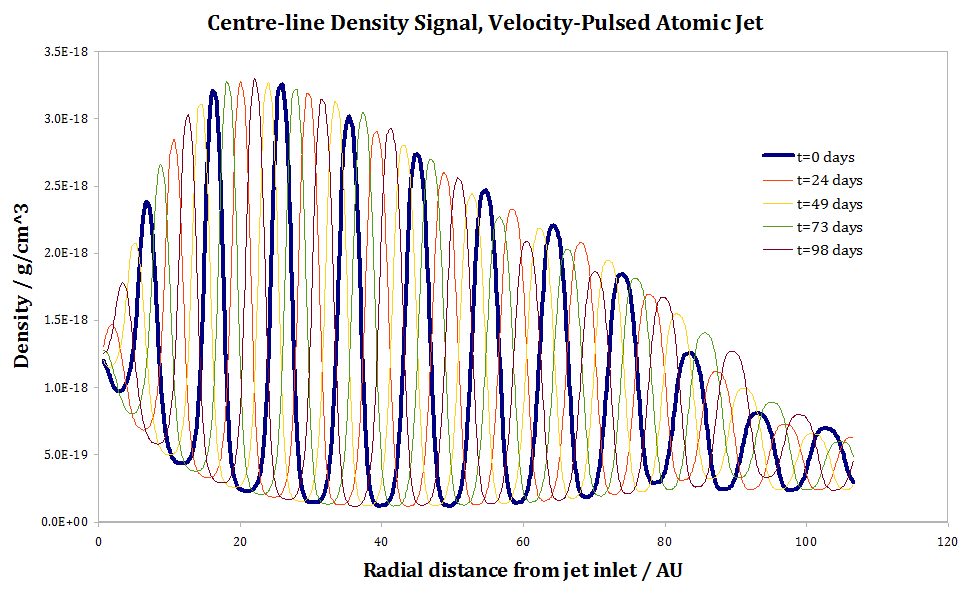}
\caption[Centre-line density signal along the axis of the pulsed atomic jets]{Centre-line density signal along the axis of the pulsed atomic jet in the dual atomic-molecular, Circumbinary Scenario s067.3.2.  An additional four time frames are included to show signal propagation.}
\label{densitycentre}
\end{figure}

The density along the axis is shown in Fig\,  \ref{densitycentre}  at five  times each separated by 24.5 days. The behaviour follows that expected for a pulsed molecular jet \citep{1997A&A...320..325S}, showing that the pulses expand once the material has been swept up into the compressed knots and the internal shocks weaken. The proper motions of the density peaks remain very close to the average jet speed of 275 km\,s$^{-1}$.

\begin{figure}
\centering
\subfloat{\includegraphics[width=1.0\linewidth, angle=0]{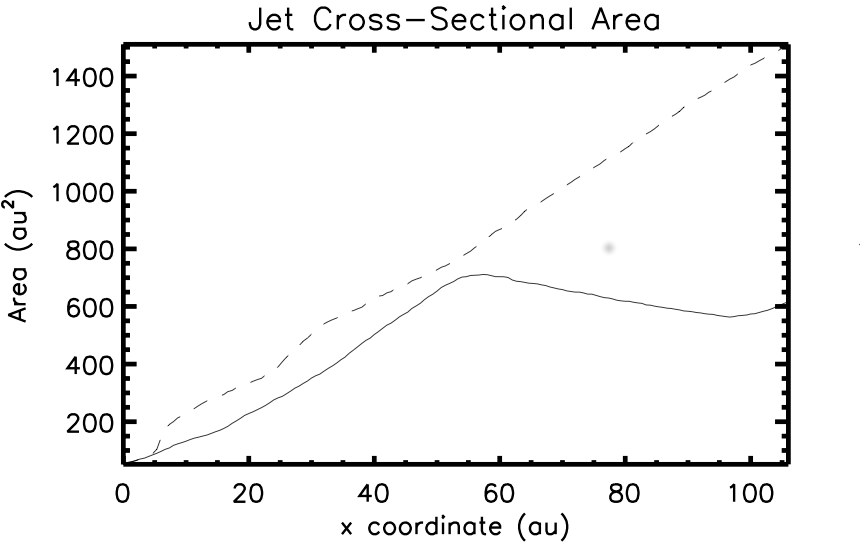}}
\hspace*{0.1\linewidth}\vspace{0em}\caption[Circumbinary Models: cross-section and position]{Circumbinary Scenario: atomic jet cross-sectional area and distance along the propagation axis at the time 87.5 years.  Model numbers:  s067.3.2 (solid line, with molecular outflow) and s067.3.3 (dashed line, without).}
\label{6732_jetbend_misc}
\end{figure}

The recollimation shoulder is the most striking new effect of the molecular outflow. As shown in
Figure \ref{6732_jetbend_misc}, at a distance x $\approx$ 55\,AU, in the presence of the surrounding molecular outflow, the jet column and hot cocoon recollimates, with the cross-sectional area falling off and then stabilising.

\begin{figure}
\centering
\subfloat{\includegraphics[width=1.0\linewidth, angle=0]{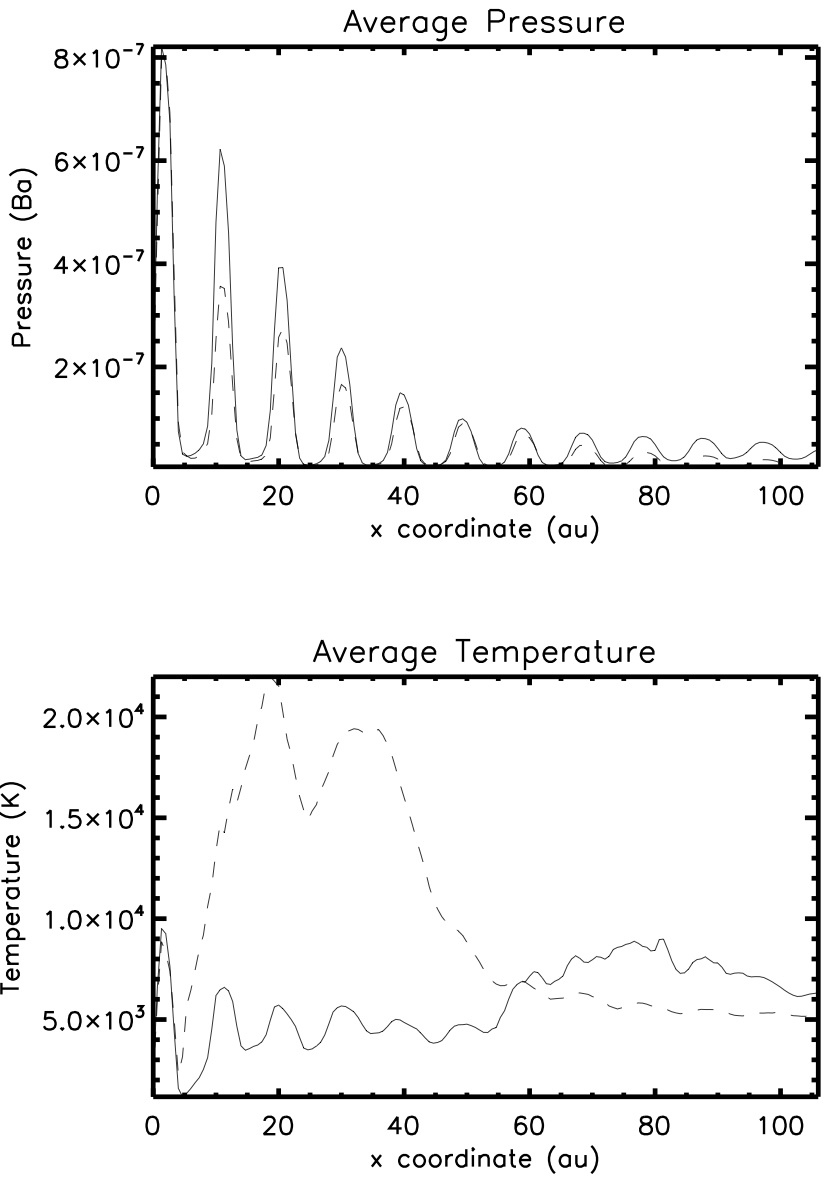}}\\[-5ex]
\hspace*{0.1\linewidth}\vspace{0em}\caption[Circumbinary Models: thermal behaviour]{Circumbinary Scenario: atomic jet pressure and temperature along the propagation axis at the time 87.5 years.  Model numbers:  s067.3.2 (solid line, with molecular outflow) and s067.3.3 (dashed line, without).}
\label{6732_jetbend_thermal}
\end{figure}

At distances x $\lesssim$ 55\,AU, we see from Figure \ref{6732_jetbend_thermal} that the ionisation and temperature are lower in the dual outflow circumbinary model than in the single outflow model, but from 55\,AU onwards this trend reverses as the ionisation in s067.3.2 dramatically rises. Thus, the molecular outflow confines the ambient medium, maintaining a high pressure which restricts 
the atomic jet after the initial rapid expansion. 

\section{Synthetic Imaging and maps}  
\label{synthetic}

\begin{figure*}
\subfloat{\includegraphics[width=0.48\linewidth]{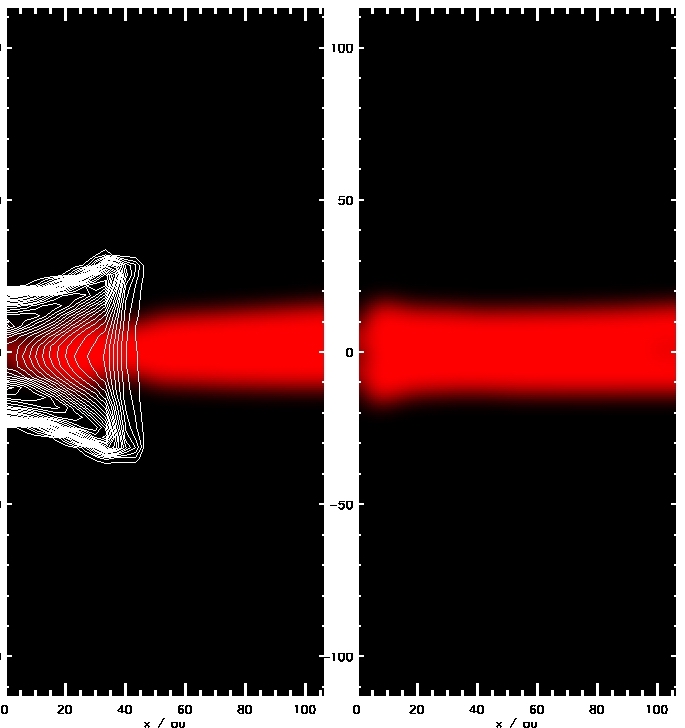}} \hfill \subfloat{\includegraphics[width=0.48\linewidth]{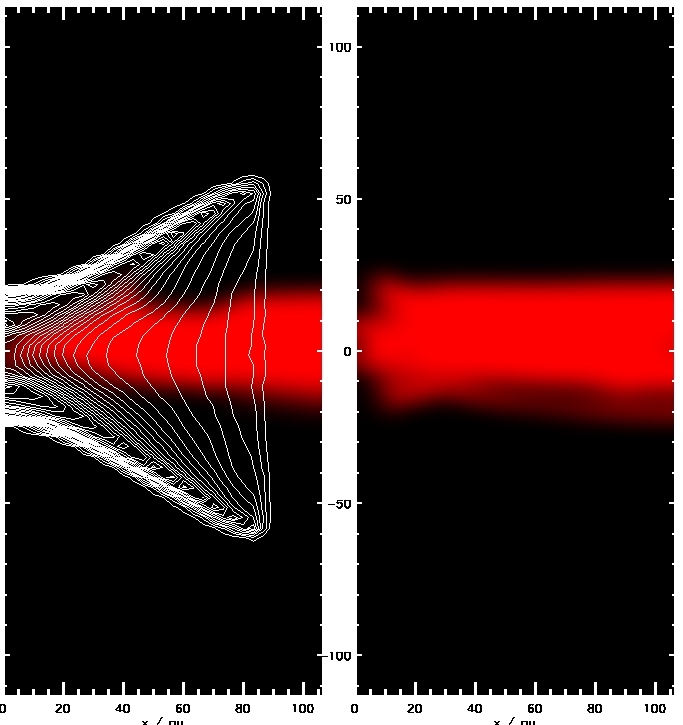}} \\  
(a)  22 years \hspace{7.8cm} (b) 44 years \\
\subfloat{\includegraphics[width=0.48\linewidth]{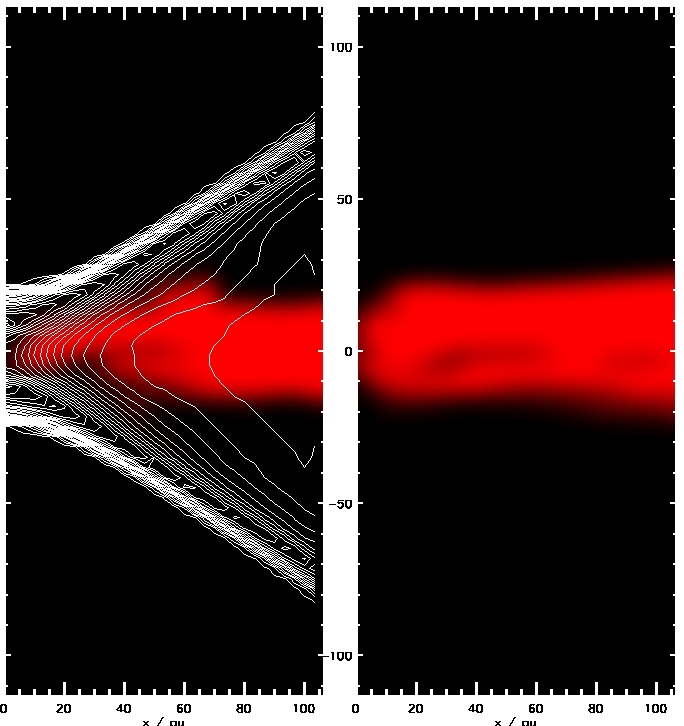}}  \hfill 
\subfloat{\includegraphics[width=0.48\linewidth]{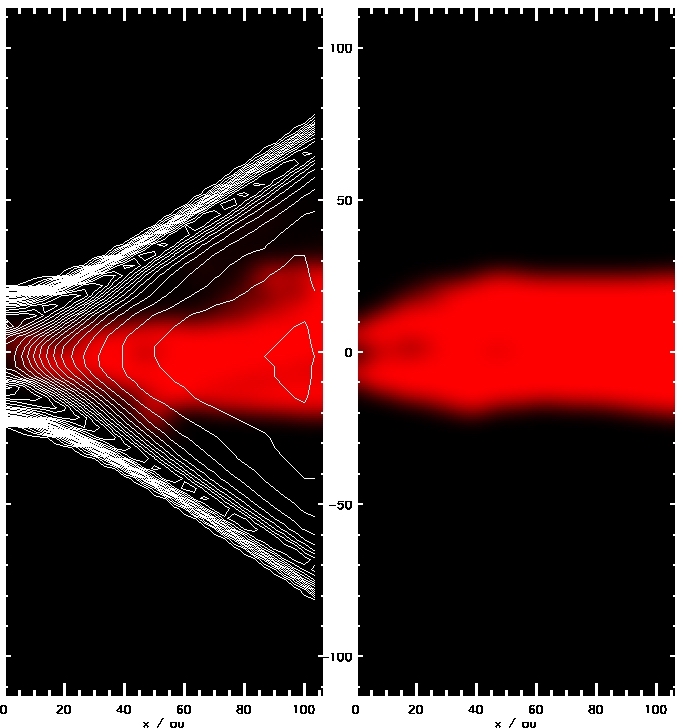}}\\
(c)  88 years \hspace{7.8cm} (d) 176 years \\
 \caption[Synthetic H-$\alpha$ Four Model Plot]
 {Synthetic H$\alpha$ images and CO contours  at the four indicated times from the Circumbinary Scenario with dual outflows (left sub-panels) and with just the atomic jet (right sub-panels)
 The optical emission is  smoothed with 14\,AU radius to match HST/WFPC2 pixel resolution of 0.1$\arcsec$ at 140\,pc.  Contours indicate molecular material (unsmoothed) from the CO J=2-1 transition.}
\label{halpha}
\end{figure*}

 Emission line properties for atomic and molecular species can be derived via post-processing of the physical variables on using statistical weights. For this purpose, we have developed an IDL code , MULTISNTH,   to generate images, position-velocity diagrams, channel maps  and mass-velocity profiles. 
In Figure\,\ref{halpha}, we present images of the H$\alpha$ 656\,nm emission superimposed on contour maps of the emission from the CO J=2-1 rotational transition at 231\.GHz.

One fascinating feature is the appearance of a wide opening angle to the H$\alpha$ jet. This occurs out to $\sim$ 15\,AU before abrupt collimation takes place (Panels (c) and (d)). This structure develops from the initial cylindrical jet structure (Panel (a)). The H$\alpha$ distribution is not completely smooth with some dark patches visible and lateral protruding bright patches. However, when the molecular flow is added, the H$\alpha$ jet is squeezed in the first 55\,AU before widening. 

The CO cone  develops at a speed of $\sim$ 10 km\,s$^{-1}$. The structure remains geometrically similar, consistent with ballistic motion. The outflow remains highly symmetric  and aligned with the jet to within 2$^\circ$: the motion of the atomic jet has a small but measurable influence on the CO. 

The H$\alpha$ jet does not display the underlying dense knots. This emission is submerged by that caused by the rapid orbit which drives oblique shock waves in the sheath and then into the jet column. This smooth appearance is at odds with the knots detected in HH\,30 which suggests that such a rapid orbiting motion of the jet source is not present.

\section{Conclusions}  
\label{conclusions}

The interaction of a jet from one component of a young binary with the heavy outflow from a circumbinary disc has been explored.
Several important assumptions were necessary to enable the three dimensional cooling and chemistry to be reasonably resolved.
Firstly, the molecular outflow is assumed to be accelerated from a region of the inner disc while the atomic jet is generated from  the primary star which orbits on a period of one year. 

The outflow axes are assumed to be aligned and the magnetic field is assumed to be dynamically negligible on scales exceeding 10\,AU. 
Thus, the  hydrodynamic simulations belong to scales exceeding 10\,AU whereas the magnetic-field driven launch  could be confined to within 0.1\,AU. Nevertheless, the field could remain dynamically  important on the 10 -- 100\,AU scale.
The initial ambient medium was taken to be  uniform and atomic; the modelled outflows were run for sufficient simulation time to fill the domain. We restrict the model to the inner 107\,AU in order to  incorporate the wide-angle molecular flow within the limits of available computing resources.

The main result derives from the rapid orbiting nozzzle from which the atomic jet is fired. The jet channel orbits rapidly, deflecting off the ambient medium, creating a sheath of ionised high-pressure gas.  The jet diverges with oblique  shocks creating a warm jet that is significantly wider than the original nozzle. The jet begins to recollimate after expanding to $\sim$ 15\,--\,20\, AU at a distance of $\sim$\,15\,AU.

In the presence of the molecular outflow, trapped material aids in the confinement of the jet which then recollimates further downstream - at 55\,AU under  the conditions taken here. In all the conditions, however, the velocity pulses in the jet do not lead to detectable knots in the H$\alpha$ emission. Instead, the ionisation, caused by the oblique shocks, overwhelms the internal  structure in the jet core.

Given these characteristics of the atomic jet, we should exclude the present Circumbinary Scenario as an interpretation of HH\,30. To work, we require a binary with a period roughly ten times longer in order for the jet dynamical time to far exceed the orbital time. That implies that the inner edge of the wind-generating circumbinary disc is also situated at a larger distance where the rotation speed is far less. It is unlikely that molecular outflow speeds of order 10\,km\,s$^{-1}$ would then result. 

On the other hand,  the Co-orbital Scenario faces similar problems, as discussed in Paper\,1. In that work, it was shown that the molecular outflow inevitably disrupts  the atomic jet.

A resolution would involve a scenario with the present circumbinary molecule flow.
The properties of the CO molecular outflow generated in this Circumbinary Scenario are consistent with the parameter set derived from the latest ALMA data \citep{2018A&A...618A.120L}.
However, to simultaneously avoid the high-excitation atomic sheath and highlight the jet knots, the atomic jet must be generated with less agitation. That requires a low-speed orbiting body which could easily be achieved  by reconsideration of the binary parameters. 
A systematic parameter study is necessary to elucidate this.

Some well-studied jets, including HH\,30,  show relatively high ionisation and wide opening angles at their base
\citep{1996ApJ...468L.103R,2000ApJ...537L..49B,2002ApJ...564..834W}. To also produce visible knots could be achieved with 
higher amplitude velocity pulsations combined with pressure pulses within the jet. Or, if the primary star  is more massive, then its orbiting path will be reduced, leading to less disruption and a diminished ionised sheath.

The present model may be  relevant to interpret the HL\,Tau jet which displays a high ionised sheath with a low excitation spine \citep{2007A&A...470..605M}. This is observed on scales of order 1,000\,AU and the triggering outflow may be attributed to an outflow from  nearby XZ\,Tau rather than an HL\,Tau wind. In addition, it should be noted that there is no evidence that HL\,Tau is a binary.

The present choice of initial conditions has a strong bearing on the results. It is clear that the high speed and neutral atomic component will lead to enhanced high ionisation regions if the jet is not uniform. Here, the orbital motion of the jet source creates a disturbed sheath which channels energy from the bulk flow into thermal modes. However, the fraction of energy dissipated per unit length and the amount then channelled into ionisation rather than kinetic energy are important factors. The potential degeneracy in the parameter space that this causes will include the orbital period with shorter periods generating narrower sheaths but dissipation over a shorter jet length.

The fact that stars commonly form not only as members of binaries or of higher order systems but often as close binaries has long been established \citep{1997A&A...321..220B}. The stars can be at distinct evolutionary stages
and can be as close as assumed in these simulations \citep{2003ApJ...583..334H}. 

To conclude,  the wide molecular flow from the circumbinary disc  disturbs  the orbiting atomic jet by squeezing the outer sheath of ionised gas. This is seen here to dominate the ionisation structure and the H$\alpha$ emission making the knots harder to distinguish.
 The orbiting atomic jet is itself very different from a fixed atomic jet as it scrapes a much wider channel around a fast spine, generating the high excitation sheath. 
The excitation level within atomic microjets will no doubt be explored 
 in the coming years, led by
the James Webb Space Telescope, the Giant Magellan Telescope and  the European Extremely Large Telescope. These are  capable of resolving structure on the scales diagnosed by these simulations.

\section*{Acknowledgements}  
\label{acks}
We  thank SEPnet and the University of Portsmouth for supplying infrastructure. 
  We also thank Simon Glover for providing support with the code.

\bibliography{thesis}

\appendix
\label{appendix}

\label{lastpage}

\end{document}